\documentclass[a4paper,twocolumn,11pt,accepted=2024-03-22]{quantumarticle}
\pdfoutput=1
\usepackage[utf8]{inputenc}
\usepackage[english]{babel}
\usepackage[T1]{fontenc}
\usepackage{amsmath}
\usepackage{hyperref}

\usepackage{tikz}
\usepackage{amssymb}
\usepackage{graphicx}
\usepackage{floatflt,epsfig} 
\usepackage{mathtools}
\usepackage{color}
\usepackage{braket}
\usepackage[normalem]{ulem}
\usepackage{blindtext}
\usepackage{lipsum}  
\usepackage{amsthm}
\usepackage{tabularx}
\usepackage{bm}
\usepackage{dsfont}
\usepackage{bbold}
\usepackage{ulem}
\usepackage{mwe}
\usepackage[percent]{overpic}
\usepackage{algorithm}
\usepackage{multirow}
\usepackage{algpseudocode}
\usepackage{csquotes}
\usepackage{bbold}
\usepackage{soul}
\usepackage{listings}
\usepackage[numbers]{natbib}

\newcommand{\rmi}{ {\ensuremath{\rm i}}}
\newcommand{\figref}[1]{\mbox{Fig.~\ref{#1}}}\newcommand{\tabref}[1]{\mbox{Table~\ref{#1}}}
\newcommand{\secref}[1]{\mbox{Section~\ref{#1}}}

\renewcommand{\eqref}[1]{\mbox{Eq.~(\ref{#1})}}
\newcommand{\e}{\ensuremath{\mathrm{e}}}
\newcommand{\bang}{qBang}
\newcommand{\qbroyden}{qBroyden}
\newcommand{\etal}{\emph{et al.}}

\newcommand{\twonorm}[1]{\ensuremath{\lVert#1\rVert}}
\newcommand{\mbf}{\mathbf}

\newcommand{\bigO}[1]{\ensuremath{\mathcal{O}(#1)}}

\begin{document}

    \title{Optimizing Variational Quantum Algorithms with qBang: Efficiently Interweaving Metric and Momentum to Navigate Flat Energy Landscapes}

	\author{David Fitzek}
	\email{davidfi@chalmers.se}
	\affiliation{Department of Microtechnology and Nanoscience, MC2, Chalmers University of Technology, 412\,96 Gothenburg, Sweden}
	\affiliation{Volvo Group Trucks Technology, 405\,08 Gothenburg, Sweden}
    \orcid{0000-0003-4268-5485}
	
	\author{Robert S. Jonsson}
	\email{robejons@chalmers.se}
	\affiliation{Department of Microtechnology and Nanoscience, MC2, Chalmers University of Technology, 412\,96 Gothenburg, Sweden}
	\affiliation{Future Technologies, Saab Surveillance, 412\,76	Gothenburg, Sweden}
    \orcid{0000-0002-8235-3058}
 
	\author{Werner Dobrautz}
	\email{werner.dobrautz@gmail.com}
	\affiliation{Department of Chemistry and Chemical Engineering, 
		Chalmers University of Technology, 412\,96 Gothenburg, Sweden}
	\orcid{0000-0001-6479-1874}
 
	\author{Christian Sch\"afer}
	\email{christian.schaefer.physics@gmail.com}
	\affiliation{Department of Microtechnology and Nanoscience, MC2, Chalmers University of Technology, 412\,96 Gothenburg, Sweden}
	\affiliation{Department of Physics, Chalmers University of Technology, 412\,96 Gothenburg, Sweden}
        \orcid{0000-0002-8557-733X}
	
\maketitle

\begin{abstract}
    Variational quantum algorithms (VQAs) represent a promising approach to utilizing current quantum computing infrastructures. VQAs are based on a parameterized quantum circuit optimized in a closed loop via a classical algorithm. This hybrid approach reduces the quantum processing unit load but comes at the cost of a classical optimization that can feature a flat energy landscape. Existing optimization techniques, including either imaginary time-propagation, natural gradient, or momentum-based approaches, are promising candidates but place either a significant burden on the quantum device or suffer frequently from slow convergence.
    In this work, we propose the quantum Broyden adaptive natural gradient (\bang{}) approach, a novel optimizer that aims to distill the best aspects of existing approaches. 
    By employing the Broyden approach to approximate updates in the Fisher information matrix and combining it with a momentum-based algorithm, \bang{} reduces quantum-resource requirements while performing better than more resource-demanding alternatives. 
    Benchmarks for the barren plateau, quantum chemistry, and the max-cut problem demonstrate an overall stable performance with a clear improvement over existing techniques in the case of flat (but not exponentially flat) optimization landscapes. \bang{} introduces a new development strategy for gradient-based VQAs with a plethora of possible improvements.
\end{abstract}

	\section{Introduction}
	
	Fostered by its anticipated potential, recent technological progress, and a surge of widespread interest, quantum computing is approaching the next level of popularity. Despite its impressive progress over the past years~\cite{Cerezo2021Variationalquantumalgorithms, Bharti2022Noisyintermediatescalequantum, Tilly2022VariationalQuantumEigensolver, Arute2019Quantumsupremacyusing, Bruzewicz2019Trappedionquantumcomputing} much remains to be accomplished before a practical use moves into reach~\cite{Daley2022Practicalquantumadvantage, Bravyi2022}.
	Two of the most severe constraints are the limited number of qubits and short coherence times~\cite{Preskill2018QuantumcomputingNISQ}. In order to combat those challenges, mixed quantum-classical algorithms, labeled variational quantum algorithms (VQAs)~\cite{Cerezo2021Variationalquantumalgorithms, Peruzzo2014, Wecker2015, McClean2016, Endo2021, Bharti2022Noisyintermediatescalequantum, Tilly2022VariationalQuantumEigensolver}, have been devised. VQAs split an optimization task into two entwined steps: (i) an energy estimation using the quantum processing unit (QPU) and (ii) a classical optimization of the characterizing parameters. Due to the existing challenges, the aim of developing VQAs is to ensure convergence while limiting the number of function evaluations on the QPU to a minimum.
	
	Classical optimizers have come a long way, from vanilla gradient descent, over natural gradient methods to the modern widely used adaptive gradient-based methods (Adam)~\cite{Kingma2017AdamMethodStochastic}.
	Similar gradient-based approaches have been introduced for quantum algorithms~\cite{Mitarai2018Quantumcircuitlearning, Banchi2021MeasuringAnalyticGradients, Schuld2019Evaluatinganalyticgradients}.
	The nature of quantum mechanics implies that, as the system size grows,  the associated Hilbert space grows exponentially.
	While it is our goal to leverage this complexity, the majority of available eigenstates are closely packed in energy, mimicking de facto thermal behavior for a local operator according to the eigenstate thermalization hypothesis~\cite{d2016quantum}.  
	Consequently, gradients, which result in small local changes in a high-dimensional Hilbert space, decrease exponentially with increasing system size, a feature known as a  barren plateau, making 
	parametrized quantum circuits (PQCs)
	prone to poor convergence.
	Albeit not directly mitigating BPs~\cite{McClean2018Barrenplateausquantum, Holmes2022, Cerezo2021Costfunctiondependent, Wang2021}, higher-order derivative information can aid in maneuvering the optimization landscape by accounting for its local curvature or metric~\cite{Stokes2020QuantumNaturalGradient, Gacon2021Simultaneous}.
	A quantity related to local curvature is the quantum Fisher information matrix (QFIM),  which appears also in the context of multi-parameter estimation~\cite{Liu2020QuantumFisherinformation}.
	
	
	Estimating gradients and higher-order derivatives of quantum circuits is, unfortunately, costly, and requires many function evaluations. Given its quadratic form, for $n_{\boldsymbol{\theta}}$ parameters the QFIM requires $\mathcal{O}({n_{\boldsymbol\theta}}^2)$ function evaluations which, considering the cost of measurements, renders its use for relevant problems challenging. Stokes \etal{}~\cite{Stokes2020QuantumNaturalGradient} introduced for pure quantum states the quantum natural gradient (QNG). 
	Block-diagonal approximations of the latter require only a linear amount of function calls but discard essential information about parameter correlation which severely limits its performance~\cite{Wierichs2020Avoidinglocalminima}.
	Generalizations of QNG to non-unitary circuits~\cite{Koczor2022Quantumnaturalgradient} as well as alternative approximation strategies have been proposed~\cite{Beckey2022Variationalquantumalgorithm, Gacon2023VariationalQuantumTime}. 
	While the specific cost of estimating the QFIM depends on the specific problem at hand, the cost for performing $\mathcal{O}({n_{\boldsymbol\theta}}^2)$ evaluations is particularly prohibitive in systems that feature vanishing gradients due to a quickly rising number of variables (e.g., the BP circuit~\cite{McClean2018Barrenplateausquantum}).
	Practical use of VQAs requires the availability of optimization strategies that provide reliable predictions with as few as possible evaluations on the QPU.
	
	In this work, we introduce the quantum Broyden adaptive natural gradient (\bang{}) approach -- an optimization strategy that augments the reliable momentum-based optimization Adam with an efficient update of the local metric based on the QFIM using the Broyden method~\cite{BROYDEN1970}. After initialization, \bang{} requires only $\mathcal{O}({n_{\boldsymbol\theta}})$ evaluations and, yet, shows considerable performance gain over QNG, Adam, and even quantum imaginary time evolution (QITE)~\cite{Motta2020Determiningeigenstatesthermal,  McArdle2019Variationalansatzbasedquantum, Yuan2019Theoryvariationalquantum, Cao2022Quantumimaginarytime} on flat optimization landscapes.
	
	The remainder of this article is structured as follows: 
	Section~\ref{sec:variational_quantum_algorithms} recapitulates VQAs, comprising the quantum approximate optimization algorithm (QAOA) and the variational quantum eigensolver (VQE), followed by a brief review of gradient-based optimization paradigms in Sec.~\ref{sec:paradigms}. Sec.~\ref{sec:brodyen_adaptive_natural_gradient} subsequently introduces the newly developed 
	\bang{} algorithm which is extensively benchmarked and discussed in Sec.~\ref{sec:results} for BP, max-cut, and quantum chemical systems. We finally conclude the discussion in Sec.~\ref{sec:conclusion} and provide an outlook toward possible applications, improvements, and future challenges.

	\section{Theory}
	
	\subsection{Variational quantum algorithms} \label{sec:variational_quantum_algorithms}
	
	VQAs are a collection of practically applicable algorithms that harness the computational capabilities of programmable quantum devices~\cite{Cerezo2021Variationalquantumalgorithms, Peruzzo2014, Yuan2019Theoryvariationalquantum}. These algorithms are well suited for the hardware constraints imposed by the current generation of quantum computers, namely short coherence times, noisy operations, and the limited number of qubits~\cite{Preskill2018QuantumcomputingNISQ}. These near-term algorithms have been proposed for a wide range of applications, including quantum chemistry~\cite{Tilly2022VariationalQuantumEigensolver}, classical optimization~\cite{Bharti2022Noisyintermediatescalequantum} and machine learning~\cite{Cerezo2021Variationalquantumalgorithms, Havlicek2019Supervisedlearningquantumenhanced}.
	
\begin{figure}
		\centering
		\includegraphics[width=0.5\textwidth]{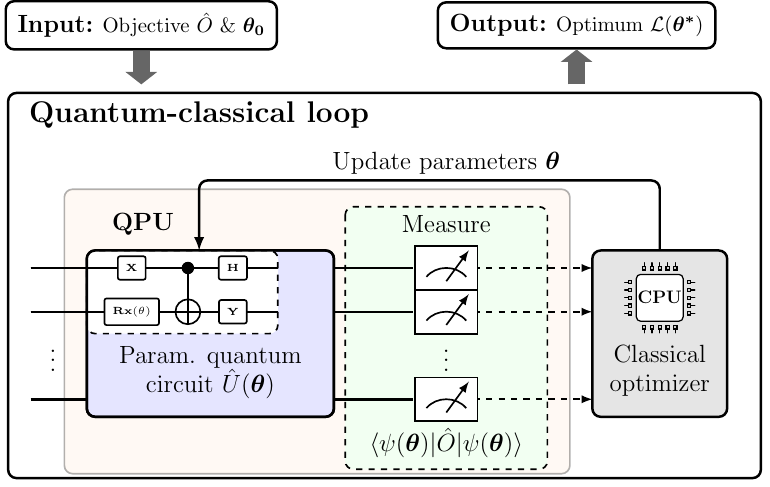}
		\caption{\label{fig:vqa}A diagrammatic representation of a VQA consists of three main elements: an objective function that defines the problem to be solved, a PQC $\hat U(\boldsymbol{\theta})$ in which parameters $\boldsymbol{\theta}$ are adjusted to minimize the objective and a classical optimizer that performs this minimization. The inputs for a VQA are the circuit Ansatz and initial parameter $\boldsymbol{\theta}_0$ values, while the outputs are the optimized parameter values $\boldsymbol{\theta}^*$ and the minimum value of the objective function, $\bra{\psi(\boldsymbol{\theta})} \hat{O}\ket{\psi(\boldsymbol{\theta})}$. } 
		\label{fig:vqa_algorithm}
	\end{figure}

	VQAs are composed of three key elements, which are represented in \figref{fig:vqa_algorithm}.
	The first component is the objective/cost function to be minimized. In our work, the cost function is expressed as the expectation value of the Hamiltonian, 
		\begin{equation} \label{eq:cost}
			\mathcal{L}(\boldsymbol{\theta}) = \bra{\psi(\boldsymbol{\theta})} \hat H \ket{\psi(\boldsymbol{\theta})},
		\end{equation}
	and provides information about the energy of the ground state of the Hamiltonian $\hat H$.
	Depending on the complexity of the Hamiltonian, different Pauli strings have to be measured to get an accurate estimate of the energy. The state
	$\ket{\psi(\boldsymbol{\theta})}$ is represented by a parametrized quantum circuit, and the optimizable parameters of the circuit are denoted as $\boldsymbol{\theta}=(\theta_1,\theta_2,\ldots,\theta_{n_\theta})^\top$.
	These parameters commonly represent the angles of unitary rotation operators. The Hamiltonian is composed of quantum operators that encode information about a chemical or classical system, such as a molecule or an optimization problem. 
	The second component is the problem-specific circuit ansatz, $\ket{\psi(\boldsymbol{\theta})}$. These ansätze are tailored to the specific problem, and numerous works focus on finding optimal PQCs~\cite{Kandala2017HardwareefficientVariationalQuantum, Farhi2014QuantumApproximateOptimization, Sim2019}.
	A shared aspect is the use of only unitary operations, a limitation that will become relevant in the subsequent sections.
	The final component is the classical optimizer, which is used to find parameters that minimize the objective function~\cite{Bharti2022Noisyintermediatescalequantum, Wierichs2022Generalparametershiftrules, Koczor2022Quantumnaturalgradient, Stokes2020QuantumNaturalGradient}.
	
	The task of VQAs is to optimize the cost function, \eqref{eq:cost}, by adjusting the tunable parameters $\boldsymbol{\theta}$ of the circuit ansatz in a closed loop. This is done by iterating between evaluating the cost function on the quantum computer and updating the parameters using a classical optimizer. The objective is to find the set of parameters, $\boldsymbol{\theta}^*$, that minimizes the cost function and provides a solution to the problem at hand. The process of evaluating the cost function and updating the parameters is repeated until the cost function converges to its minimum value or a stopping criterion is met.
	Current limitations in the available complexity of circuits are thus circumvented by dividing the optimization problem into small sets of quantum evaluations steered via classical parameter optimization.
	The circuit ansatz, cost function, and classical optimizer are problem-specific, and the choice of these components can significantly affect the algorithm's performance. 
	
	VQAs offer a versatile framework that can be broadly categorized into several areas of application. While QAOA~\cite{Farhi2014QuantumApproximateOptimization} is often employed for classical optimization problems and VQE~\cite{Peruzzo2014, Kandala2017HardwareefficientVariationalQuantum, Tilly2022VariationalQuantumEigensolver} is commonly used for solving quantum eigenvalue problems, these categories are not exhaustive.
	
	QAOA has been proposed to solve various classical optimization problems~\cite{Farhi2014QuantumApproximateOptimization,Lucas2014Isingformulationsmany, Cerezo2021Variationalquantumalgorithms, Hadfield2019quantumapproximateoptimization,Svensson2021HeuristicMethodsolve} and is a candidate for hybrid quantum-classical computation. Here, optimization problems are encoded into an Ising Hamiltonian~\cite{Lucas2014Isingformulationsmany}.
	QAOA typically suggests a circuit ansatz $\ket{\psi (\boldsymbol{\theta})}$ composed of the consecutive application of two non-commuting operators. One operator encodes the optimization problem and the other serves as a mixing Hamiltonian.
	The goal is to optimize the parameters, $\boldsymbol{\theta}$, of the quantum circuit to minimize $\mathcal{L}(\boldsymbol{\theta})$, and thereby find the solution to the optimization problem. 
	Once the quantum circuit has been optimized, bitstrings are sampled to obtain approximate solutions to the classical optimization problem.
	
	In contrast, the VQE is the most widely studied quantum algorithm to minimize a given cost function, usually the energy, of a given quantum system, \eqref{eq:cost}. A prominent example is the solution of Schr\"odinger's equation for molecular systems. 
	A selected PQC is initialized, and the corresponding energy of the output state is subsequently evaluated on a quantum computer.
	Information about energy, gradients, and the metric can be inferred from multiple evaluations of the circuit and then used
	to update the parameters of the circuit with classical optimization methods~\cite{Lavrijsen2020ClassicalOptimizersNoisy}.
	This process is repeated until the expectation value converges to the ground-state energy of the system (see \secref{sec:variational_quantum_algorithms}).
	The VQE algorithm has been applied in various fields, including quantum chemistry~\cite{Cao2019} and materials science~\cite{Lordi2021}.
	
	\subsection{Existing optimization paradigms}\label{sec:paradigms}
	
	Here we review the existing optimization paradigms that inspire the \bang{} approach.
	\subsubsection{Gradient-based Optimization}
	
	A vital component of every variational algorithm is the classical optimizer. Here, the task of the classical computer is to iterate the parameters from an initial guess $\boldsymbol{\theta}_0$ such that the cost, \eqref{eq:cost}, is minimized. Generally, this requires several iterations, depending on the quality of the initial guess. Assuming the cost function is differentiable, this procedure can be realized with \textit{gradient descent} (GD).  GD uses the parameter update rule $   \boldsymbol{\theta}_{k+1} = \boldsymbol{\theta}_{k} - \eta \nabla \mathcal{L}_k $, where the step size $\eta\in \mathbb{R}^+$ controls how much each iteration is allowed to change the parameters and $\nabla \mathcal{L}_k\equiv \nabla\mathcal{L}(\boldsymbol{\theta}_k)$ is the gradient of the cost function at iteration $k$. The norm of the gradient $\twonorm{\nabla \mathcal{L}_k}_2$ can be used as a  criterion to determine when to stop the GD algorithm, as a zero norm gradient implies a stationary point.
	Gradients of quantum circuits can be obtained via finite-difference methods, linear combination of unitaries~\cite{Schuld2019Evaluatinganalyticgradients} and without the need for additional hardware by evaluating the cost function at two shifted parameter positions and using the rescaled difference of the results as an unbiased estimate of the derivative~\cite{Crooks2019Gradientsparameterizedquantum, Wierichs2022Generalparametershiftrules, Schuld2019Evaluatinganalyticgradients}.
	
	GD-based methods have apparent limitations. If the cost function is relatively flat, the gradient will be small, and the GD may require unfeasibly many iterations to converge, even on ideal quantum devices. The noisy results on realistic devices put additional strain on the optimizer to escape flat energy landscapes as quickly as possible.
	As long as the cost function gradients are not completely vanishing, this problem may be mitigated by the extension of GD to include higher-order derivatives.
	For a second-order algorithm, this introduces the Hessian $\mathbf{H}$ and results in the Newton method, $\boldsymbol{\theta}_{k+1}=\boldsymbol{\theta}_k-\eta \mathbf{H}_k^{-1}\nabla \mathcal{L}_k$. However, these higher-order methods are not always applicable, as the Hessian may not be positive semi-definite~\cite{Martens2020, Martens2012}. Additionally, computing the Hessian is computationally expensive if the parameter space is large. To overcome these challenges, there are several quasi-Newton methods that can efficiently estimate the Hessian, such as the Broyden-Fletcher-Goldfarb-Shanno (BFGS) algorithm or the Gauss-Newton method~\cite{Shanno1970, BROYDEN1970, Fletcher1970, Goldfarb1970}. 
	
	Other methods exist that are tailored to navigate flat energy landscapes. For an intuitive picture, consider a ball rolling down in a frictionless bowl. Instead of stopping at the bottom, the accumulated momentum pushes it forward and keeps the ball rolling back and forth. This idea is used in momentum-based optimizers, illustrated in a simplified form
	\begin{align} 
		\boldsymbol{m}_k &= \beta \boldsymbol{m}_{k-1}+(1-\beta) \nabla \mathcal{L}_k \\
		\boldsymbol{\theta}_{k+1} &= \boldsymbol{\theta}_{k}- \eta \boldsymbol{m}_k,
	\end{align}
	where each step is a linear combination of the previous update and the current gradient, with $\boldsymbol{m}_k$ being the momentum accumulated during the optimization process, $\beta$ the decay rate and $\eta$ the step size.
	Compared to GD, these methods are more effective at escaping local minima~\cite{Ruder2017overviewgradientdescent}.
	The Adam~\cite{Kingma2017AdamMethodStochastic} momentum-based optimizer is widely used throughout different scientific disciplines and has proven versatile and consistent in performance.
	
	The optimization of VQAs can suffer when the energy surface becomes flat. To handle this issue, two directions can be taken. One approach is to find a good initial state that can be easily obtained and prepared on the quantum device. A typical example for chemistry applications is the uncorrelated Hartree–Fock state, but it can be expected that more complex systems will require correlated initial states. In the second approach, we utilize information about the local metric to guide each step toward the minimum, which will be discussed in detail in the following section. Overall, finding practical solutions to this problem is crucial for successfully implementing VQAs.
	
	\subsubsection{\label{sec:qite} Metric-informed Optimization: quantum imaginary time evolution and quantum natural gradient}

	As stated above, VQAs rely on a parametrization of the wave function in which the parameters represent phases of unitary gates acting on an input state. A small change in a parameter $\delta\theta_i$ not only results in changes in the observable of interest, as utilized by GD, but also in the associated metric $\langle \psi (\delta\theta_j)\vert \psi(\delta\theta_i)\rangle$. This additional information can provide a more suitable direction for the optimization trajectory. We will briefly review QITE and QNG, representing the two most widely discussed metric-informed optimization strategies.
	
	QITE~\cite{Motta2020Determiningeigenstatesthermal,  McArdle2019Variationalansatzbasedquantum, Yuan2019Theoryvariationalquantum, Cao2022Quantumimaginarytime} is based on the \enquote{Wick-rotated}($\tau = \rmi t$)~\cite{Wick1954}   imaginary time Schr\"odinger equation
	\begin{equation}\label{eq:imag}
		\begin{split}
		\frac{\partial \ket{\Psi(\tau)}}{\partial\tau}&= -\hat H \ket{\Psi(\tau)} \\
			\text{or } 
			\ket{\Psi(\tau + \Delta\tau)} &= N(\tau)^{-1} \e^{-\Delta\tau \hat H}\ket{\Psi(\tau)},  \\
			 \text{with } N(\tau) &  = \sqrt{\bra{\Psi(\tau)} \e^{-2\Delta\tau\hat H} \ket{\Psi(\tau)}}
		\end{split}
	\end{equation}
	and is a quantum algorithm to find the ground and excited states~\cite{Tsuchimochi2023} of a quantum system. It is a variant of the imaginary time evolution (ITE) algorithm~\cite{vonderLinden1992, Ceperley1995, Trivedi1990, Guther2020}, which is a well-established technique in \enquote{classical} computational physics for finding the ground state of a system. The iterative application of the exponential operator with sufficiently small time-steps $\Delta\tau$~\cite{Trivedi1990} is exponentially damping higher energy contributions, resulting in a convergence to the ground state $\ket{\Psi_0}$ if the initial state $\ket{\Psi(0)}$ has a non-zero overlap with the ground state~\cite{Motta2020Determiningeigenstatesthermal, McArdle2019Variationalansatzbasedquantum}.
	However, since $\e^{-\Delta\tau \hat H}$ is not unitary, it 
	is not straightforward to directly implement ITE on quantum hardware. 
	One option, which we will pursue in this work, is to cast QITE into a hybrid quantum-classical variational form (VarQITE)~\cite{McArdle2019Variationalansatzbasedquantum, Yuan2019Theoryvariationalquantum} (Fig.~\ref{fig:vqa}),
	where the target state $\ket{\Psi(\tau)}$  is encoded by a PQC 
	$\ket{\psi(\boldsymbol{\theta}(\tau))} = \hat U(\boldsymbol{\theta}(\tau))\ket{\psi_0}$
	and the time-evolution is mapped to the parameters $\boldsymbol{\theta}(\tau)$ of the variational ansatz.
	The rule to update the parameters $\boldsymbol{\theta}_k$ for the next iteration $k+1$ at (imaginary) time $\tau + \Delta\tau$ is
	obtained by applying McLachlan's variational principle~\cite{McLachlan1964} to \eqref{eq:imag}, 
	minimizing the difference of the time evolution of the ansatz state $\ket{\psi(\tau)}\equiv\ket{\psi(\boldsymbol{\theta}(\tau))}$ to the exact imaginary time evolution
	\begin{equation}\label{eq:mclachlan}
		\delta \twonorm{\left(\partial/\partial \tau + \hat{H} - E_{\tau}\right)\ket{\psi(\tau)}}_2 = 0,
	\end{equation}
	where $\twonorm{\ket{\psi}}_2 = \sqrt{\braket{\psi\vert\psi}}$ is the 2-norm of a quantum state $\ket{\psi}$ and $E_{\tau} = \bra{\psi(\tau)}\hat H\ket{\psi(\tau)}$ is the expected energy at time $\tau$. 
	Solving \eqref{eq:mclachlan} yields the imaginary-time derivative of the parameters
	\begin{equation}\label{eq:lse}
		\frac{\partial \boldsymbol{\theta}}{\partial\tau} = -2\;\mbf F^{-1} \nabla \mathcal{L}, 
	\end{equation}
	where $\mathbf{F}$ is the QFIM and  $\nabla \mathcal{L}$ the cost gradient. \eqref{eq:lse} allows updating the parameters for the next iteration, i.e., with a fixed time-step $\Delta\tau$ and the Euler method
	\begin{equation}
		\boldsymbol{\theta}_{k+1} = \boldsymbol{\theta}_k + \Delta\tau  \frac{\partial \boldsymbol{\theta}}{\partial\tau}=  \boldsymbol{\theta}_k - \frac{\Delta\tau}{2} \mbf F^{-1}_{k} \nabla \mathcal{L}_k,
	\end{equation}
	or higher-order methods~\cite{Zoufal2021}.
	$\Delta \tau$ is equivalent to a step size, $\eta$, in the above mentioned GD update rule.
	The elements of the QFIM are given by
	\begin{equation}
		\mathbf{F}_{ij} = 4 \mathrm{Re}\left[ \left\langle\partial_{\theta_i} \psi \vert \partial_{\theta_j} \psi\right\rangle
		-  \left\langle\partial_{\theta_i}\psi\vert\psi\right\rangle \left\langle\psi \vert\partial_{\theta_j}\psi\right\rangle \right], \label{eq:fisher}
	\end{equation}
	where, $\partial_{\theta_i} \equiv \frac{\partial}{\partial_{\theta_i}}$. 
	There is a close relation between the QFIM and the Fubini-Study metric, which is the metric of parametrized pure quantum states $\ket{\psi}$, see the Supplemental Information (SI) \secref{app:connectionFIMHessian} and Refs. numbers~\cite{Fubini1908, Study1905, Yao2022, Wilczek1989, Hackl2020, zhoujiang2019, Liu2020QuantumFisherinformation, Giovannetti2011, Petz1996} for details. 
	The QFIM $\mbf F$ encodes the nontrivial geometry of the parameter space~\cite{Provost1980, Petz1996} and is the quantum-analog of the classical Fisher information matrix, which is the unique Riemannian metric associated to a probability density function~\cite{Park2020Geometrylearningneural, Braunstein1994, Facchi2010}. 
	
	QNG~\cite{Stokes2020QuantumNaturalGradient} is another metric-informed optimization technique based on the principles of natural gradient descent by Amari \emph{et al.}~\cite{Amari1996,Amari1998NaturalGradientWorks, Amari1998b, Amari2000AdaptiveMethodRealizing,Park2020Geometrylearningneural}, initially developed for optimizing neural networks. As VarQITE, the natural gradient considers the geometry of the function's parameter space and is calculated using the inverse of the QFIM~\cite{Liu2020QuantumFisherinformation, Meyer2021}.
	Thus, using the QNG results in steps that are more aligned with the geometry of the parameter space and allows for faster convergence, crossing of local minima, and helps the algorithm to escape regions with vanishing gradients~\cite{Stokes2020QuantumNaturalGradient,Wierichs2020Avoidinglocalminima,Gacon2021Simultaneous,Huembeli2021Characterizinglosslandscape,McClean2018Barrenplateausquantum,Grant2019,Cerezo2021Costfunctiondependent}.
	VarQITE and QNG are equivalent when the energy of the system,  $E = \braket{\hat H}$, is used as the cost function~\cite{Amari1998NaturalGradientWorks, Martens2020, Stokes2020QuantumNaturalGradient, Koczor2022Quantumnaturalgradient}, as considered in this work, see \eqref{eq:cost}.
	
	The major drawback of QITE and QNG is that computing the entire QFIM for an ansatz with
	$n_{\boldsymbol{\theta}}$ parameters is computationally expensive and requires measuring $\bigO{n_{\boldsymbol{\theta}}^2}$
	terms every iteration.
	Existing approximations such as the (block-) diagonal approximation of Stokes 
	\etal{}~\cite{Stokes2020QuantumNaturalGradient} reduce the scaling to 
	linear in the number of parameters, but discarding the off-diagonal elements omits essential information about correlation within the system and leads to an overall suboptimal performance~\cite{Wierichs2020Avoidinglocalminima}.
	
	The metric $\mbf F$ and the gradient $\mbf \nabla \mathcal{L}$ can be directly evaluated on quantum hardware~\cite{Schuld2019Evaluatinganalyticgradients, Yuan2019Theoryvariationalquantum, Sokolov2022, Dobrautz2023}. 
	It should be noted that the metric is frequently singular due to over-parametrization of the chosen circuit ansatz and requires regularization~\cite{Stokes2020QuantumNaturalGradient, Wierichs2020Avoidinglocalminima} or comparable strategies~\cite{Bromley2020,Zoufal2021}.

	\subsection{Quantum Broyden Adaptive Natural Gradient}\label{sec:brodyen_adaptive_natural_gradient}
	In this section, we introduce 
	\bang{}, that combines the Broyden  quasi-Newton method with the natural gradient and adaptive momentum approaches. 
	We discuss the core components of 
	\bang{}, as well as its motivation, mechanics, and resources required on the programmable quantum device.
	We also introduce a simplified version of our optimization approach, which we refer to as \qbroyden{}.
	
	The algorithms  \bang{} and \qbroyden{} utilize an adaptive approach to approximate the QFIM, drawing inspiration from the works of Amari, Park, and Fukumizu~\cite{Amari2000AdaptiveMethodRealizing, Park2000Adaptivenaturalgradient}. The intuition behind this approach can be understood as follows. We would like to retain the benefits of the natural gradient method without computing the QFIM at each iteration. For this reason, we assume that the QFIM varies slowly as the parameter space is traversed. For time step $k$, we use a metric denoted by the matrix $\mathbf{B}_k$. Between steps, the metric is updated with a rank-1 perturbation given by the current gradient.
	 In particular, $\mathbf{B}_{k+1}$ is realized as  a low-pass filter process with learning rate $\varepsilon_k$, allowing the metric to pick up momentum as the parameter space is traversed, given by the relation
	\begin{align}
		\mathbf{B}_{k+1} = (1-\varepsilon_k)\mathbf{B}_{k}+\varepsilon_k\nabla \mathcal{L}_k\nabla \mathcal{L}_k^\top.\label{eq:matrixfilter}
	\end{align}
	Conceptually, this updates the local metric with an approximation of the Hessian. 
	In the classical setting, the Hessian is equivalent to the Fisher information matrix for certain classes of optimisation problems, e.g., with Gaussian statistics or if the connection between the probability of encountering a given state decreases exponentially with its energy density (see SI  \secref{app:connectionFIMHessian}).
	More generally, the connection to curvature is also found in the equivalence between the \textit{classical} Fisher information matrix and the Hessian of the relative entropy between two parametrically separated distributions~\cite{amari2016information}.
	We want to note that recently, Dash \emph{et al.}~\cite{Dash2024} have related the QFIM with the Hessian in the context of neural quantum states by using the infidelity with respect to the exact ground state as the cost function.
	The famous BFGS algorithm uses similar ideas as \eqref{eq:matrixfilter} but differs in approximating the Hessian using \textit{two} rank-1 updates. 
		
Instead of updating and then inverting  $\mathbf{B}_{k+1}$, we utilise the Sherman-Morrison formula to equivalently perform the update on the inverse as
	\begin{align}
		\mathbf{B}_{k+1}^{-1} 
		=& \left[\mathbb{1} - \frac{\varepsilon_k\mathbf{B}_k^{-1}\nabla \mathcal{L}_k\nabla \mathcal{L}_k^\top}{1-\varepsilon_k(1-\nabla \mathcal{L}_k^\top \mathbf{B}_k^{-1} \nabla \mathcal{L}_k)}\right] \frac{ \mathbf{B}_k^{-1}}{1-\varepsilon_k}.\label{eq:updaterule} 
	\end{align}
	We select the hyperparameter $\varepsilon_k$ according to a decaying filter $\varepsilon_k = \varepsilon_0/(k+1)$~\cite{Amari1998NaturalGradientWorks}. 
	
	Algorithm~\ref{alg:bang} presents the pseudo-code of the \bang{} optimizer, which will be briefly exercised in the following. The algorithm takes as input the learning rates $\eta=0.01$ and $\varepsilon_0=0.2$, the decay rates $\beta_1=0.9$ and $\beta_2=0.999$, the convergence criterion $\gamma$, and the PQC $U(\boldsymbol{\theta})$ with the initial parameter vector $\boldsymbol{\theta}_0\in \mathbb{R}^{n_{\theta}}$.
	In the initialization step, the algorithm sets the iteration counter $k\leftarrow 0$, the momentum vector $\boldsymbol{m}_{-1} \leftarrow \mathbf{0}$ and the biased variance vector $\boldsymbol{v}_{-1} \leftarrow \mathbf{0}$, whose role will become apparent in the following. 
	The matrix $\mathbf{B}_{0}$ is initialized using, either, the full Fisher information matrix ($\mathbf{F}$) or an approximation as introduced in~\cite{Stokes2020QuantumNaturalGradient}. 
	Other choices for the matrix $\mathbf{B}_{0}$ would result in variations of the algorithm.
	The optimization starts with the estimation of the cost function $\mathcal{L}(\boldsymbol{\theta}_k)$ and its gradient $\nabla \mathcal{L}(\boldsymbol{\theta}_k)$ through quantum circuits, followed by the update of the momentum and variance vectors, similar to the Adam algorithm~\cite{Kingma2017AdamMethodStochastic}.
	Specifically, the algorithm calculates a weighted average of past gradients $\boldsymbol{m}_k$, with the weight given by a parameter $\beta_{1}$, and uses this as a moving direction.
	It incorporates a moving average of the squared gradient, $\boldsymbol{v}_k \leftarrow \beta_{2}  \boldsymbol{v}_{k-1}+\left(1-\beta_{2}\right) \nabla \mathcal{L}(\boldsymbol{\theta}_k) \odot \nabla \mathcal{L}(\boldsymbol{\theta}_k)$, with the weight given by a second parameter $\beta_{2}$. The vector $\boldsymbol{v}_k$ can be interpreted as the variance under the assumption of a vanishing average. Its magnitude provides information about the reliability of a gradient estimate. The moving averages are then adjusted for bias via division with $(1-\beta^{k+1}_{(1/2)})$, delivering $\boldsymbol{\widehat{m}}_k$ and $\boldsymbol{\widehat{v}}_k$. The variance vector $\boldsymbol{\widehat{v}}_k$ is then used to rescale the effective momenta into a sliding trust region $\{\boldsymbol{p}_k\}_l\leftarrow \{\boldsymbol{\widehat{m}}_k \}_l/\left(\sqrt{\{\boldsymbol{\widehat{v}}_k \}_l}+\kappa\right),\  \forall l\in \{1,2,\ldots, p\}$, i.e., increasing the stability of the algorithm by shortening unreliable steps.
	Unless the convergence criterion is reached,
	the algorithm updates the parameter vector and the  metric based on the update rule~\eqref{eq:updaterule}. It also rescales $\varepsilon_k$ with the learning rate schedule, resulting in smaller updates with increasing number of optimization steps. 
	Otherwise, if the convergence criterion is satisfied, the algorithm stops the iteration and outputs the optimal parameter vector $\boldsymbol{\theta}^*$.
	We suggest reinitializing \bang{} once the update of the Fisher information matrix becomes minute, which might appear for particularly long optimization trajectories but has not been encountered in this work.
	
	Algorithm~\ref{alg:qbroyden} presents a simplified version of our optimization approach, which we refer to as \qbroyden{}. Unlike \bang{}, \qbroyden{} does not incorporate momentum and variance update rules and instead utilizes only the  metric to update the parameter vector at each optimization step. Consequently, \qbroyden{} is more closely related to QNG and VarQITE than \bang{}. 
	
	Our framework surrounding \eqref{eq:updaterule} has several advantages. Firstly, the Fisher information matrix is guaranteed to be positive semi-definite~\cite{Liu2020QuantumFisherinformation}. 
	With the Gauss-Newton-like update, we maintain the positive semi-definiteness property through the optimisation, see SI \secref{app:connection_of_varqite_and_qng} and Martens \textit{et al.}~\cite{Martens2020}. In fact, we apply a small regularisation to the initial QFIM to ensure that $\mathbf{B}_0$ is positive definite. This is an important feature since it can happen that the QFIM is singular, particularly in overparameterized systems with multiple layers. Additionally, because the  QFIM is not recalculated at each time step, this framework significantly reduces  the necessary number of circuit evaluations. Lastly, incorporating momentum updates not only results in superior speed but also increases the stability with respect to hyperparameter changes (illustrated in Sec.~\ref{sec:hyperparameter_resilience}).
	
	We want to note that a potential drawback of approximating the QFIM is that the resulting algorithms technically lose theoretically ensured convergence properties of QITE~\cite{Motta2020Determiningeigenstatesthermal, McArdle2019Variationalansatzbasedquantum}.
	However, this was not an issue for all the problems studied in this work. On the contrary, qBang ensured a faster and more stable convergence. 

	Regarding circuit evaluations, our proposed method reduces cost and increases efficiency.
	Each optimization step requires $ \mathcal{O}( n_{\boldsymbol{\theta}})$ circuit evaluations, which is on par with Adam due to the parameter-shift rule~\cite{Wierichs2022Generalparametershiftrules, Crooks2019Gradientsparameterizedquantum}. 
	QNG without any approximation scales as $\mathcal{O}(n_{\boldsymbol{\theta}}^2)$ due to estimating the full Fisher information matrix~\cite{Meyer2021}.
	Our proposed optimizers, \bang{} and \qbroyden{}, require as many circuit evaluations in the first step as QNG, and only $\mathcal{O}(n_{\boldsymbol{\theta}})$ circuit evaluations per subsequent optimization step.
	The following sections demonstrate that the most striking advantage of \bang{} is its efficiency. 
	
	\begin{algorithm}[H]
		\caption{\bang{} }
		\begin{algorithmic}[1]
			\State  \textbf{Input:} learning rates $\eta=0.01$, $\varepsilon_0=0.2$
			\State  \textbf{Input:} decay rates $\beta_1=0.9$ and $\beta_2=0.999$
			\State  \textbf{Input:} convergence criterion $\gamma$
			\State  \textbf{Input:} PQC $U(\boldsymbol{\theta})$
			\State \textbf{Input:} Initial parameter vector $\boldsymbol{\theta}_0\in \mathbb{R}^{n_\theta}$.
		\State \textbf{Initialization:}   $k\leftarrow0$, $\boldsymbol{m}_{-1} \leftarrow \mathbf{0}$, $\boldsymbol{v}_{-1} \leftarrow \mathbf{0}$, $\mathbf{B}_{0}^{-1}$ via QNG, QFIM or Identity 
		\State \texttt{not\_converged}$\leftarrow$\texttt{true} 
		\While{\texttt{not\_converged}}     
		\State \textbf{QC}: estimate $\mathcal{L}(\boldsymbol{\theta}_k)$
		\State \textbf{QC}: estimate $\nabla \mathcal{L}(\boldsymbol{\theta}_k)$
		\State $\boldsymbol{m}_k\leftarrow\beta_{1} \boldsymbol{m}_{k-1}+\left(1-\beta_{1}\right) \nabla \mathcal{L}(\boldsymbol{\theta}_k)$ 
		\State $\boldsymbol{v}_k \leftarrow \beta_{2}  \boldsymbol{v}_{k-1}+\left(1-\beta_{2}\right)  \nabla \mathcal{L}(\boldsymbol{\theta}_k) \odot \nabla \mathcal{L}(\boldsymbol{\theta}_k)$ 
		\State $\boldsymbol{\widehat{m}}_k \leftarrow \boldsymbol{m}_k /\left(1-\beta_{1}^{k+1}\right)$
		\State $\boldsymbol{\widehat{v}}_k \leftarrow \boldsymbol{v}_k /\left(1-\beta_{2}^{k+1}\right)$
		\State $\{\boldsymbol{p}_k\}_l\leftarrow \{\boldsymbol{\widehat{m}}_k \}_l/\left(\sqrt{\{\boldsymbol{\widehat{v}}_k \}_l}+\kappa\right),\  \forall l\in \{1,2,\ldots, p\}$ 
		\If{$\|\mathbf{B}_k^{-1}\boldsymbol{p}_k\|_2>\gamma$}
		\State $\boldsymbol{\theta}_{k+1} \leftarrow \boldsymbol{\theta}_{k} - \eta \, \mathbf{B}_k^{-1}\boldsymbol{p}_k / ((k + 2) - 1 )^{\epsilon_0}$ 
		\State $\varepsilon_k\leftarrow \frac{\varepsilon_0}{k+1}$        
		\State $\mathbf{B}_{k+1}^{-1}\leftarrow$  \eqref{eq:updaterule}
		\State $k\leftarrow k+1$
		\Else     
		\State \texttt{not\_converged}$\leftarrow $\texttt{false} 
		\State      $\boldsymbol{\theta}^*\leftarrow\underset{\{\boldsymbol{\theta}_n\}_0^k }{\operatorname{argmin}} \,\mathcal{L}(\boldsymbol{\theta}_n)$              
		\EndIf
		\EndWhile
		\State{\Return $\boldsymbol{\theta}^*$}
	\end{algorithmic}
	\label{alg:bang}
\end{algorithm}

\begin{algorithm}[H]
	\caption{\qbroyden{} }
	\begin{algorithmic}[1]
		\State  \textbf{Input:} learning rates $\eta=0.01$, $\varepsilon_0=0.2$
		\State  \textbf{Input:} convergence criterion $\gamma$
		\State  \textbf{Input:} PQC $U(\boldsymbol{\theta})$ 
		\State \textbf{Input:} Initial parameter vector $\boldsymbol{\theta}_0\in \mathbb{R}^{n_\theta}$.
	\State \textbf{Initialization:}   $k\leftarrow0$, $\mathbf{B}_{0}^{-1}$ via QNG, QFIM or Identity 
	\State \texttt{not\_converged}$\leftarrow$\texttt{true} 
	\While{\texttt{not\_converged}}     
	\State \textbf{QC}: estimate $\mathcal{L}(\boldsymbol{\theta}_k)$
	\State \textbf{QC}: estimate $\nabla \mathcal{L}(\boldsymbol{\theta}_k)$
	\If{$\|\mathbf{B}_k^{-1} \nabla \mathcal{L}(\boldsymbol{\theta}_k)\|_2>\gamma$}
	\State $\boldsymbol{\theta}_{k+1} \leftarrow \boldsymbol{\theta}_{k} - \eta \, \mathbf{B}_k^{-1} \nabla \mathcal{L}(\boldsymbol{\theta}_k) $ 
	\State $\varepsilon_k\leftarrow \frac{\varepsilon_0}{k+1}$        
	\State $\mathbf{B}_{k+1}^{-1}\leftarrow$  \eqref{eq:updaterule}
	\State $k\leftarrow k+1$
	\Else     
	\State \texttt{not\_converged}$\leftarrow $\texttt{false} 
	\State      $\boldsymbol{\theta}^*\leftarrow\underset{\{\boldsymbol{\theta}_n\}_0^k }{\operatorname{argmin}} \,\mathcal{L}(\boldsymbol{\theta}_n)$              
	\EndIf
	\EndWhile
	\State{\Return $\boldsymbol{\theta}^*$}
\end{algorithmic}
\label{alg:qbroyden}
\end{algorithm}

\section{Results}\label{sec:results}

This section presents numerical results from noise-free simulations of the new optimizers applied to several important classes of problems. We focus only on hybrid quantum-classical algorithms, which combine quantum and classical processing.
The necessary quantum circuits for this study are available on GitHub~\cite{fitzek2023qflow} and additional information is provided in the SI.

Considering that quantum circuit queries are costly, our main goal is to reduce the number of circuit evaluations to obtain the parameters encoding the ground state of the PQC. Therefore, the key metric is the number of circuit evaluations.
See Section~\ref{sec:brodyen_adaptive_natural_gradient} for the scaling of the number of circuit evaluations for each optimizer.
Another important metric to assess the performance of the optimization is the approximation ratio. It describes how close the energy of the optimized quantum circuit is to the ground state energy. Formally, the approximation ratio is defined as
\begin{equation}
r = \frac{E_{\rm opt} - E_{\rm max}}{E_{\rm min} - E_{\rm max}} \label{eq:approximation_ratio},
\end{equation}
where $E_{\rm opt}$ is the energy obtained after optimization, and $E_{\rm min}$ and $E_{\rm max}$ are the theoretical minimum and maximum energy values, respectively. 

We compare the optimizers Adam~\cite{Kingma2017AdamMethodStochastic}, QNG~\cite{Stokes2020QuantumNaturalGradient} with the block-diagonal approximation, as well as \qbroyden{} and \bang{} using either the full or block-diagonal Fisher information in the first iteration. 
We largely exclude VarQITE in the following due to its prohibitive cost but show results for individual trajectories in SI \secref{app:single_trajectories}. It should be noted that the computational overhead for VarQITE might reduce in relation to gradient estimates when using advanced sampling techniques~\cite{VanStraaten2021MeasurementCostMetricAware}. 
However, the cost of simulation with sampling is considerably larger than the here employed state propagation.
For QNG and VarQITE, in case the QFIM is singular, we employ a Tikhonov regularization~\cite{Tikhonov1995} and add $10^{-7}$ to its diagonal. 
Both algorithms of \qbroyden{} and \bang{} use an initial filter parameter of $\epsilon_0=0.2$. For QNG and Adam, we use default parameters provided in~\cite{Bergholm2022PennyLaneAutomaticdifferentiation}.

We use identical step sizes for all algorithms to ensure a fair comparison but  emphasize that the optimal step size will depend on the problem and algorithm at hand. Our investigation is comprehensive, accounting for statistical features in the random initialization, but not exhaustive, given the infinite combinations of hyperparameters and VQAs.

\subsection{Barren plateau circuit}\label{sec:barren_plateau_circuit}

We start by illustrating the performance of the newly proposed optimizers on the BP circuit introduced in Ref.~\cite{McClean2018Barrenplateausquantum}. This quantum circuit was initially designed to show that highly expressible circuits come with a caveat, i.e., the more freedom we give a quantum circuit, the more difficult the optimization due to vanishing gradients in the exponentially growing Hilbert space~\cite{Cerezo2021Costfunctiondependent}. The consequence: simple gradient-based optimizers fail.

\begin{figure*}
\centering
\includegraphics[width=\linewidth]{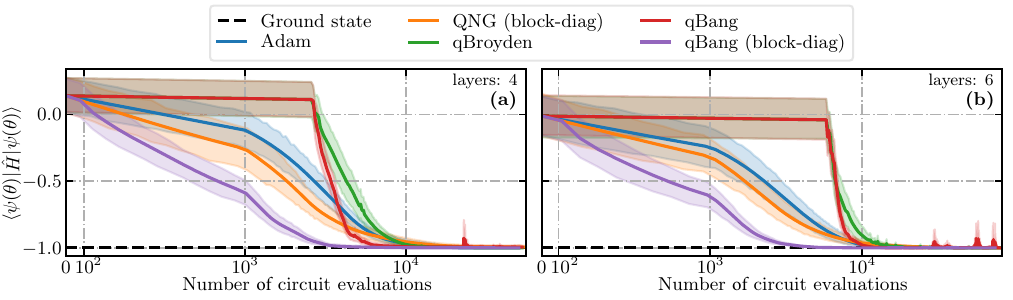}
\caption{Comparison of optimization performance of Adam, QNG, \qbroyden{}, and \bang{} in finding the ground state of the BP circuit. $\langle \psi(\boldsymbol{\theta}) | \hat H | \psi(\boldsymbol{\theta}) \rangle$ is shown as a function of the  number of circuit evaluations. The step size is fixed at $\eta=0.01$, and the results are averaged over $25$ random initializations of parameters. The PQC used consists of $4$ and $6$ layers as depicted in subplots (a) and (b), respectively. The initial plateau in the optimization using \qbroyden{} and \bang{} arises from the significant cost of initially measuring the QFIM.
}
\label{fig:barrenplateau_average}
\end{figure*}

Our circuit consists of an initial fixed layer of $R_y(\pi/4)$ gates acting on $9$ qubits, followed by $l$ layers of parameterized Pauli rotations with an entangling layer of controlled-$Z$ gates. The objective operator is $\hat H=\hat Z_1\hat Z_2$ with a ground state energy of $-1$. 
The relative quality of the optimization will depend on the initial configuration, i.e., drawing a meaningful conclusion for the performance of an optimizer for a given problem requires a statistical analysis. In this manuscript, we obtain the expectation value $\langle \psi(\boldsymbol{\theta}) | \hat{H} | \psi(\boldsymbol{\theta}) \rangle$ for a parametrization of the wavefunction which is to be optimized.
Our plots show  the mean and variance of $25$ trajectories with randomly initialized parameters (the same for all algorithms) and a step size of $\eta=0.01$. The PQC considered has $4$, $6$, $8$, and $10$ layers, respectively. Figure~\ref{fig:barrenplateau_average} illustrates the performance as a function of circuit evaluations using $4$ and $6$ layers. 

The QNG (block-diagonal) optimizer shows a moderate improvement over Adam within the initial 5000 evaluations for a small set of parameters but loses this initial advantage in the long run. 
\bang{}, on the other hand, is substantially faster.
Approximating the QFIM as block-diagonal reduces the computational cost for the first iteration and explains the reduction in the required number of evaluations for the convergence of \bang{} (block-diag). 
The early plateau observed in the performance of \qbroyden{} and \bang{} results from the upfront computational effort needed to estimate the QFIM.
More relevant in practice is the number of circuit evaluations required to approximate the ground state accurately. To evaluate this, we determine the number of circuit evaluations necessary to reach an approximation ratio of $0.99$ and present the results in \tabref{tab:barrenplateau_circuit_evaluations}.
As shown in the table, \bang{} (block-diag) substantially outperforms Adam and QNG, requiring merely a third of the circuit evaluations.

While the BP circuit is of no practical use, it illustrates that \bang{} is a highly competitive optimizer when handling almost flat energy surfaces. We will briefly discuss classical optimization problems before moving on to quantum chemistry, arguably the most promising application for quantum computing to this date.

\begin{table}
\caption{Comparison of the number of circuit evaluations required for four optimizers to reach  an approximation ratio of $r=0.99$ for the BP circuit, with the results averaged over the $25$ optimization trajectories. The PQC used range from $4$, $6$, $8$, to $10$ layers. 
\enquote{bd} indicates the block-diagonal approximation. 
}
\label{tab:barrenplateau_circuit_evaluations} \small
\begin{tabular}{lrrrr}
	\hline
	& \multicolumn{4}{c}{Layers} \\
	\cline{2-5}
	Optimizer                  &  4         & 6                & 8                & 10  \\
	\hline
	Adam                    & 10700         & 10300       & 10200             & 13000             \\
	\bang{}                 & 5980          & 9750        & 16900             & 25300               \\
	\bang{}  (bd)    & \textbf{3290} & \textbf{3490} & \textbf{4150}     & \textbf{5330}               \\
	\qbroyden{}             & 10300         & 13100         & 16100             & 25300  \\
	\qbroyden{}  (bd)& 8990          & 11400         & 13800             & 17900  \\
	QNG  (bd)        & 12300         & 17300         & 18500             & 26900           \\
	\hline
\end{tabular}
\end{table}

\subsection{Quantum Approximate Optimization Algorithm} 

Classical combinatorial optimization can be just as hard as the optimization of quantum systems. QAOA represents a subclass of VQAs that handles the question if quantum computing could assist such classical combinatorial optimization.

\begin{figure*}
\centering    
\includegraphics{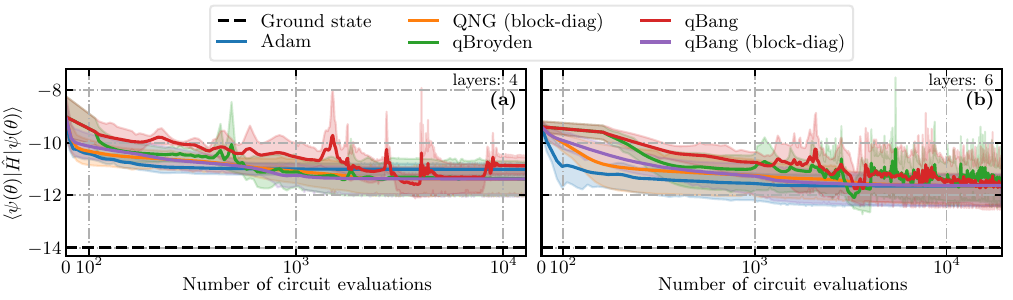}
\caption{Ground state optimization performance of Adam, QNG with block-diagonal approximation, \qbroyden{} with full Fisher information matrix, and \bang{} with full Fisher matrix and block-diagonal approximation of the QAOA circuit of an eight qubit max-cut problem instance using a PQC. The expectation value, $\langle \psi(\boldsymbol{\theta}) | \hat H | \psi(\boldsymbol{\theta}) \rangle$ is shown as a function of the  number of circuit evaluations. The step size is fixed at $\eta=0.06$, and the results are averaged over five random initializations of parameters. The PQC used consists of $4$ and $6$ layers as depicted in subplots (a) and (b), respectively.}
\label{fig:qaoa_average}
\end{figure*}

We study the max-cut problem for which the cost (or energy) of the classical problem is mapped to an Ising Hamiltonian~\cite{Lucas2014Isingformulationsmany}.
The Hamiltonian for the max-cut problem is encoded using eight qubits on the quantum device. 
The optimization performance of the different optimizers is displayed in \figref{fig:qaoa_average} against the number of circuit evaluations. The results are averaged over five random initializations of parameters and a step size of $\eta = 0.06$. We show the optimization trajectories for the $4$- and $6$-layered circuits in subplots (a) and (b), respectively. 
In \tabref{tab:qaoa_approximation_ratio}, we compare the approximation ratios for the quantum state with the lowest expectation value, obtained by averaging over five trials for $4$-, $6$-, $8$-, and $10$-layered quantum circuits. 

The optimization trajectories shown in \figref{fig:qaoa_average} are similar in convergence behavior. One notable difference is the oscillations that \qbroyden{} and \bang{} exhibit after many circuit evaluations using the full Fisher information. 
The oscillations result from incomplete updates of the off-diagonal elements in the Fisher information, which pushes the optimization away from the optimal direction. We elaborate on this feature in the SI \secref{sec:update_rule_ablation_study}.
Using the block-diagonal approximation ensures a smoother optimization. Alternatively, \qbroyden{} and \bang{} could be reinitialized whenever instabilities occur.

\tabref{tab:qaoa_approximation_ratio} shows the approximation ratio averaged over five trajectories. 
Our proposed algorithms perform well on the $4$- and $6$-layered quantum circuits, while Adam outperforms all optimizers for $8$- and $10$-layers. 
Overall we observe only minor differences in convergence behavior, and the significant deviation from the optimal solution demonstrates that QAOAs face a serious challenge.
It is important to note that the used circuit ansatz is likely incapable of representing a quantum state near the ground state of the classical optimization problem.

\begin{table}
\caption{Ground state energy approximation ratios of Adam, QNG with block-diagonal approximation, \qbroyden{}, and \bang{} with full Fisher information and block-diagonal approximation for the max-cut Ising Hamiltonian. 
	Results for PQCs with $4$, $6$, $8$, and $10$ layers are shown.
	The values are obtained from the quantum state with the expectation value closest to the ground state averaged over the five optimization pathways with a maximum length of $1100$ optimization steps.
 \enquote{bd} indicates the block-diagonal approximation. 
 }
\label{tab:qaoa_approximation_ratio}\small
\begin{tabular}{lllll} 
	\hline
	& \multicolumn{4}{c}{Layers} \\
	\cline{2-5}
	Optimizer                &  4        & 6         & 8         & 10  \\
	\hline
	Adam                    & 0.787     & 0.832    & \textbf{0.896}   & \textbf{0.91}  \\
	\bang{}                 & \textbf{0.829}& 0.866    & 0.872   & 0.832 \\
	\bang{}  (bd)        & 0.813     & 0.83     & 0.826   & 0.86 \\
	\qbroyden{}             & 0.816     & \textbf{0.89}& 0.846   & 0.879 \\
	\qbroyden{}  (bd)    & 0.814     & 0.833    & 0.881   & 0.888 \\
	QNG  (bd)            & 0.814     & 0.833    & 0.87    & 0.879 \\
	\hline
\end{tabular}
\end{table}

\subsection{Variational Quantum Eigensolver} 
Solving Schr\"odinger's equation is challenging, yet essential to understand chemistry. 
In this study, we concentrate on investigating three prototypical molecular benchmark systems: hydrogen four ($\rm{H}_4$), lithium hydride (LiH), and the water molecule ($\rm{H}_2\rm{O}$).
We employed minimal basis sets (STO-6G) for all quantum chemistry problems and used a frozen core approximation for LiH and H$_2$O (with the 1s orbital of Li and O, respectively, frozen)~\cite{Helgaker2000}. 
To construct the quantum circuits, we used the Jordan-Wigner Fermion-to-qubit mapping and employed a hardware-efficient ansatz~\cite{Kandala2017HardwareefficientVariationalQuantum} that utilizes $8$, $10$, and $12$ qubits for $\rm{H}_4$, LiH, and $\rm{H}_2\rm{O}$, respectively. This ansatz is composed of $l$ layers, each comprising a tunable $R_y(\theta)$ gate on each qubit register, followed by a closed ring of CNOT gates.
We compare the algorithm's performance with random and Hartree-Fock parameter initializations. 
Details of the molecular geometries and the Hartree-Fock parameter initialization can be found in the SI \secref{app:initialization_using_hartree_fock}.
We used \texttt{Pennylane}~\cite{Bergholm2022PennyLaneAutomaticdifferentiation} with the built-in \texttt{PySCF} interface~\cite{Sun2020} to setup our molecular systems and perform the Fermion-to-qubit mapping.

\begin{figure*}
\centering
\includegraphics[width=\textwidth]{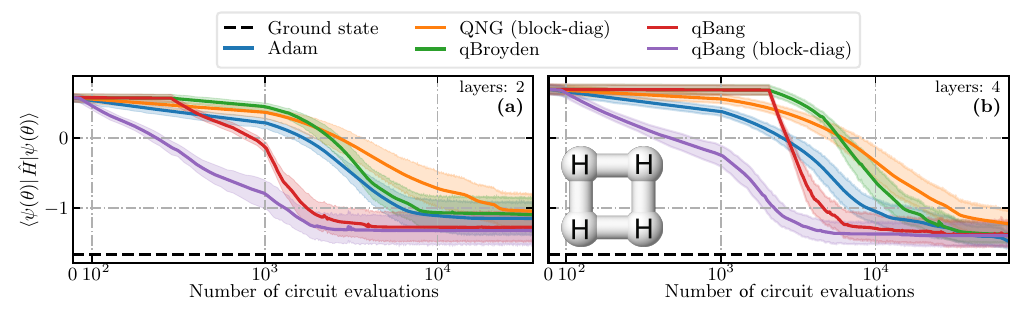}
\caption{Comparison of optimization performance of Adam, QNG using the block-diagonal approximation, \qbroyden{} using the full Fisher matrix, and \bang{} with the full Fisher information and  block-diagonal approximation in finding the ground state of $\rm{H}_4$ using a PQC. The expectation value, $\langle \psi(\boldsymbol{\theta}) | \hat H | \psi(\boldsymbol{\theta}) \rangle$ is shown as a function of the  number of circuit evaluations. The step size is fixed at $\eta=0.01$, and the results are averaged over $15$ random initialization of parameters. The PQC consists of $2$ and $4$ layers, as shown in subplots (a) and (b), respectively. The initial plateau in the optimization using \qbroyden{} and \bang{} arises from the significant cost of initially measuring the QFIM.}
\label{fig:h4_average}
\end{figure*}

Our results provide insight into the feasibility and limitations of hardware-efficient circuit ansätze for preparing the ground state of molecular systems. 
In addition to assessing the optimization performance, we also analyze the physical soundness of the quantum states generated with the lowest overall energy. 
To this end, we calculate various observables, including the particle number, $\hat{N}$, the total spin projection observable, $\hat{S}_z$, and the total spin observable, $\hat{S}^2$, based on the optimized quantum state $\ket{\psi(\boldsymbol{\theta})}$. 

\subsubsection{Hydrogen square, $\rm{H}_4$}
\label{sec:h4}
We studied four hydrogen atoms, H$_4$, arranged in a square geometry with a side length of 2.25~\AA. 
Figure~\ref{fig:h4_average} presents the mean energy as a function of the number of circuit evaluations for circuits with two and four layers. \bang{} requires substantially fewer circuit evaluations, \qbroyden{} is on par with Adam and the performance of QNG is limited. The latter is likely due to the importance of off-diagonal components in the QFIM for correlated systems.

Upon further analysis of the quantum states generated by the PQCs, we find that,  for all optimizers, the particle number $\langle\hat{N}\rangle$ and total spin projection $\langle \hat{S}_z\rangle$ observables are in proximity, but not in precise agreement with, the physical ground state (see \tabref{tab:h4_lih_table}). The deviations are most severe for the total spin $\langle \hat{S}^2\rangle$ and illustrate that the total energy is not the only observable of interest for the optimization in VQEs.
This issue is a common challenge for hardware-efficient ansätze and stems from the choice of the circuit ansatz rather than the optimization algorithm itself (see also SI \secref{app:chemistry_applications}).
We verified the numerics with an equivalent Qiskit implementation providing the same hyperparameter and initial conditions leading to the same optimization trajectory.

\begin{table*}
\caption{Converged optimization results for PQCs, representing \( \rm{H}_4 \) and \( \rm{LiH} \). Results for \( \rm{H}_4 \) are averaged over 15 optimization trajectories, while results for \( \rm{LiH} \) are averaged over 10 optimization trajectories. The ground truth for each observable is shown in the column $\langle \hat{O}\rangle_{\Psi}$. Observables are calculated for circuits with layers ranging from $1$ to $4$ based on the variational quantum state with minimum expectation value along the optimization trajectory. Bold symbols indicate the optimizer that gets closest to the ground truth. The column labeled \bang{} shows results by starting with the full Fisher information matrix, and the column to the right labeled $F^{k=0}_{\text{block-diag}}$ are results starting with the block-diagonal approximation.}
\centering
\small
\begin{tabular}{ccc|cccccc}
    \hline
    \multicolumn{8}{c}{$\rm{\bf H}_{\mathbf{4}}$} \\
    \hline
    $\hat{O}$ & $ \langle \hat{O} \rangle_{\Psi} $ & $l$ & Adam & \bang{} & $F^{k=0}_{\text{block-diag}}$ &  \qbroyden{} & QNG \\
    \hline
    \multirow{4}{*}{$ \hat{H}$} & \multirow{4}{*}{\textbf{-1.665}} & 1 & \textbf{-1.08} & -1.05 & -1.05 & -1.03 & -1.03 \\
    & & 2 & -1.15 & -1.21 & \textbf{-1.25} & -1.21 & -1.18 \\
    & & 3 & \textbf{-1.37} & -1.34 & -1.35 & -1.35 & -1.34 \\
    & & 4 & \textbf{-1.46} & -1.42 & -1.41 & -1.4 & -1.37 \\
    \hline
    \multirow{4}{*}{$\hat{N}$} & \multirow{4}{*}{\textbf{4}} & 1              & \textbf{3.8} & 3.67 & 3.62 & 3.6 & 3.59 \\
    & & 2 & \textbf{3.93} & 3.88 & 3.91 & 3.9 & 3.89 \\
    & & 3 & 3.83 & 3.84 & \textbf{3.87} & \textbf{3.87} & 3.86 \\
    & & 4 & 3.83 & \textbf{3.9} & 3.89 & 3.88 & 3.88 \\
    \hline
    \multirow{4}{*}{$\hat{S}_z$} & \multirow{4}{*}{\textbf{0}} & 1 & -0.5 & -0.5 & -0.5 & -0.43 & \textbf{-0.39} \\
    & & 2 & \textbf{-0.17} & -0.19 & -0.18 & -0.18 & -0.19  \\
    & & 3 &-0.09 & \textbf{-0.07} & -0.09 & -0.14 & -0.14 \\
    & & 4 & -0.31 & -0.17 & -0.16 & \textbf{-0.14} & \textbf{-0.14} \\
    \hline
    \multirow{4}{*}{$\hat{S}^2$} & \multirow{4}{*}{\textbf{0}} & 1 & \textbf{1.52} & 1.73 & 1.81 & 1.63 & 1.54 \\
    & & 2 & 1.48 & \textbf{1.39} & 1.4 & 1.45 & 1.46 \\
    & & 3 & 1.79 & \textbf{1.74} & 1.84 & 1.83 & 1.82 \\
    & & 4 & 1.56 & 1.49 & 1.46 & \textbf{1.42} & 1.44 \\
    
    \hline
    \multicolumn{8}{c}{$\rm{\bf LiH}$} \\
    \hline
    $\hat{O}$ & $ \langle \hat{O} \rangle_{\Psi} $ & $l$ & Adam & \bang{} & $F^{k=0}_{\text{block-diag}}$ & \qbroyden{} & QNG \\
    \hline
    \multirow{4}{*}{$ \hat{H} $} & \multirow{4}{*}{\textbf{-7.972}} & 1 & -7.33 & -7.35 & -7.35 & \textbf{-7.36} & \textbf{-7.36} \\
    & & 2 & -7.75 & -7.81 & \textbf{-7.84} & -7.82 & -7.79 \\
    & & 3 & -7.66 & -7.69 & \textbf{-7.72} & -7.67 & -7.64 \\
    & & 4 & -7.73 & -7.77 & \textbf{-7.81} & -7.77 & -7.74 \\
    \hline
    \multirow{4}{*}{$\hat{N}$} & \multirow{4}{*}{\textbf{2}} & 1 &  3.0 & 2.9 & \textbf{2.87} & 2.93 & 2.98 \\
    & & 2 & 2.2 & 2.1 & \textbf{2.07} & \textbf{2.07} & 2.09    \\
    & & 3 & 2.8 & 2.71 & \textbf{2.61} & 2.68 & 2.74 \\
    & & 4 & 2.5 & 2.5 & 2.44 & 2.44 & \textbf{2.42} \\
    \hline
    \multirow{4}{*}{$\hat{S}_z$} & \multirow{4}{*}{\textbf{0}} & 1 & 0.1 & 0.05 & \textbf{0.03} & 0.08 & 0.1 \\
    & & 2 & \textbf{-0.3} & -0.35 & -0.37 & -0.43 & -0.44\\
    & & 3 &\textbf{-0.0} & -0.01 & -0.05 & 0.07 & 0.13\\
    & & 4 & 0.09 & 0.12 & 0.06 & \textbf{0.05} & 0.06\\
    \hline
    \multirow{4}{*}{$\hat{S}^2$} & \multirow{4}{*}{\textbf{0}} & 1 & 1.65 & 1.53 & 1.48 & 1.41 & \textbf{1.37} \\
    & & 2 & 1.25 & 1.13 & \textbf{1.02} & 1.14 & 1.25 \\
    & & 3 & \textbf{1.6} & 1.82 & 1.78 & 1.77 & 1.75  \\
    & & 4 & 1.13 & 0.93 & \textbf{0.76} & 0.84 & 0.94 \\
    \hline
\end{tabular}
\label{tab:h4_lih_table}
\end{table*}

\subsubsection{Lithium hydride, $\rm{LiH}$} 

\begin{figure*}
\centering
\includegraphics[width=\textwidth]{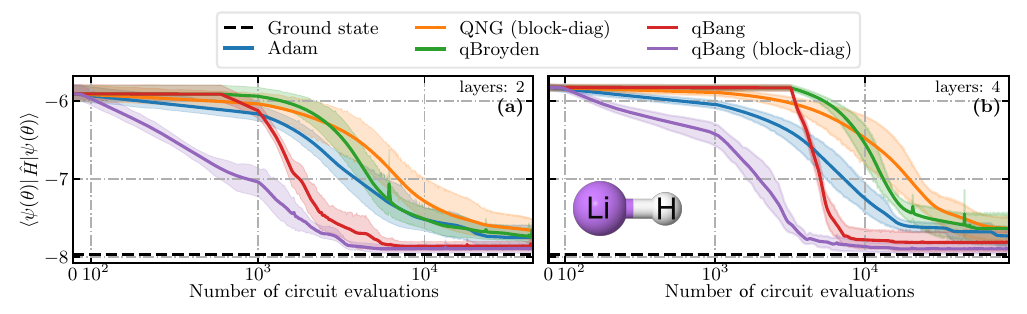}
\caption{Comparison of optimization performance for four optimizers in finding the ground state of $\rm{LiH}$ using a PQC. The optimizers evaluated are Adam, Quantum Natural Gradient using the block-diagonal approximation, \qbroyden{} using the full Fisher information matrix, and \bang{} with the full Fisher information and  block-diagonal approximation. The expectation value, $\langle \psi(\boldsymbol{\theta}) | \hat H | \psi(\boldsymbol{\theta}) \rangle$ is shown as a function of the  number of circuit evaluations. The step size is fixed at $\eta=0.01$, and the results are averaged over $5$ random initializations of parameters. The PQC used consists of $2$ and $4$ layers, as shown in subplots (a) and (b), respectively. The initial plateau in the optimization using \qbroyden{} and \bang{} arises from the significant cost of initially measuring the QFIM.}
\label{fig:lih_average}
\end{figure*}

We studied LiH at a bond distance of 1.59~\AA\; with the 1s orbital of Li frozen. 
Figure~\ref{fig:lih_average} clarifies that the conclusions drawn for $\rm{H_4}$ can be largely transferred to $\rm{LiH}$: \bang{} vastly outperforms its competitors and consistently finds the best estimation for the energy closest to the ground state. Furthermore, once the optimum has been obtained, the comparably small variance of the $10$ trajectories indicates a reliable optimization process. Consistent with $\rm{H_4}$, $\langle \hat{S}^2 \rangle$ challenges all optimizers (see \tabref{tab:h4_lih_table}). 

\subsubsection{Water, $\rm{H}_2\rm{O}$}
\begin{figure*}
\centering
\includegraphics[width=\textwidth]{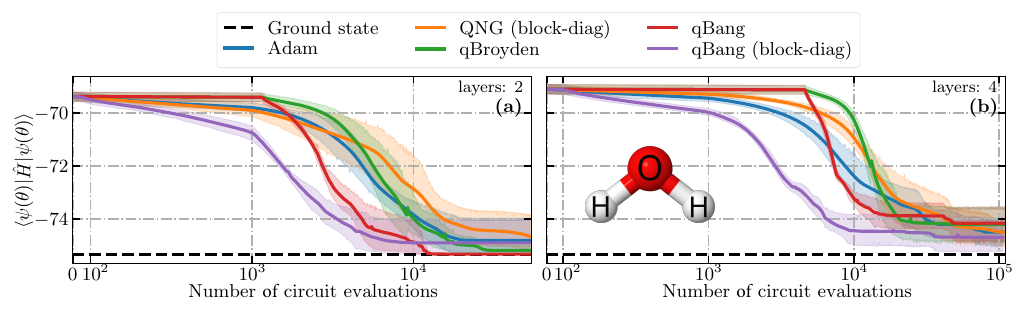}
\caption{Comparison of optimization performance for four optimizers in finding the ground state of $\rm{H}_2\rm{O}$ using a PQC. The optimizers evaluated are Adam, QNG using the block-diagonal approximation, \qbroyden{} using the full Fisher matrix, and \bang{} with the full Fisher information and  block-diagonal approximation. The expectation value, $\langle \psi(\boldsymbol{\theta}) | H | \psi(\boldsymbol{\theta}) \rangle$ is shown as a function of the  number of circuit evaluations. The step size is fixed at 0.01 and the results are averaged over $5$ random initializations of parameters. The PQC consists of $2$ and $4$ layers, as shown in subplots (a) and (b), respectively. 
	The initial plateau in the optimization using \qbroyden{} and \bang{} arises from the significant cost of initially measuring the QFIM.
}
\label{fig:h2o_average}
\end{figure*}

We studied H$_2$O with an OH distance of 0.7~\AA\; and with an $\angle$(HOH) of 104.48$^{\circ}$ with the 1s orbital of O frozen. 
Figure~\ref{fig:h2o_average} illustrates the mean expectation value as a function of the number of circuit evaluations for quantum circuits consisting of two and four layers averaged over five trials. As before, \bang{} outperforms Adam and QNG. Interestingly, \bang{} with the full Fisher information is the only optimizer that manages to discover the exact ground state energy of the system in one of the optimization trajectories for two layers.
The optimized circuits corresponding to the state with the lowest overall energy are analyzed in Table~\ref{tab:h2o_table}, showing an overall good performance of \bang{} and Adam. 

Overall, \bang{} deliver accurate results for quantum chemistry applications at a discount. An important question remains: How resilient is this observation against changes in hyper-parameters or noise?

\begin{table*}[ht]
\caption{Converged optimization results for PQCs, representing $\rm{H}_2\rm{O}$. Results are averaged over five optimization trajectories. 
	The ground truth for each observable is shown in the column $\langle \hat{O}\rangle_{\Psi}$. Observables are calculated for circuits with layers ranging from $1$ to $4$ based on the variational quantum state with minimum expectation value along the optimization trajectory. Bold symbols indicate the optimizer that gets closest to the ground truth. The column labeled \bang{} shows results by starting with the full Fisher information matrix, and the column to the right labeled $F^{k=0}_{\text{block-diag}}$ are results starting with the block-diagonal approximation.}
 \small
\begin{center}
	\begin{tabular}{ccc|cccccc}
		\hline
		$\hat{O}$ & $ \langle \hat{O} \rangle_{\Psi} $ & $l$ & Adam & BANG & $F^{k=0}_{\text{block-diag}}$ & \qbroyden{} & QNG \\
		\hline
		\multirow{4}{*}{$ \hat{H}$} & \multirow{4}{*}{\textbf{-75.36}} & 1 & \textbf{-73.45}  & -73.23 & -73.15 & -73.34 & -73.36 \\
		& & 2 & -74.82 & \textbf{-75.08}  & -75.02 & -75.07 & -75.0 \\
		& & 3 & -73.59 & -74.01 & -74.04 & -74.11 & \textbf{-74.18} \\
		& & 4 & \textbf{-74.5} & -74.34 & -74.46 & -74.4 & -74.44  \\
		\hline
		\multirow{4}{*}{$\hat{N}$} & \multirow{4}{*}{\textbf{8}}          & 1 &\textbf{7.8} & 7.3 & 7.13 & 7.35 & 7.44 \\
		& & 2 & 7.6 & \textbf{ 7.8} & \textbf{7.8} & 7.79 & 7.74   \\
		& & 3 & 7.3 & 7.55 & \textbf{7.6} & 7.58 & 7.56 \\
		& & 4 & 7.9 & \textbf{7.95} & 7.87 & 7.75 & 7.76 \\
		\hline
		\multirow{4}{*}{$\hat{S}_z$} & \multirow{4}{*}{\textbf{0}}        & 1 &  \textbf{-0.1} & -0.55 & -0.7 & -0.52 & -0.48    \\
		& & 2 & -0.2 & -0.1 & -0.1 & -0.09 & \textbf{-0.07}     \\
		& & 3 & 0.15 & 0.12 & 0.2 & 0.13 & \textbf{0.09} \\
		& & 4 &  0.05 & \textbf{0.0} & -0.04 & -0.01 & 0.02  \\
		\hline
		\multirow{4}{*}{$\hat{S}^2$} & \multirow{4}{*}{\textbf{0}} & 1 & 2.95 & 3.28 & 3.38 & 2.89 & \textbf{2.86}    \\
		& & 2 & 0.5 & \textbf{0.25} & 0.28 & 0.26 & 0.32   \\
		& & 3 & 2.63 & 1.99 & 1.9 & 1.66 & \textbf{1.53} \\
		& & 4 & 1.53 & 1.46 & 1.32 & \textbf{1.27} & 1.29 \\
		\hline
	\end{tabular}
\end{center}
\label{tab:h2o_table}
\end{table*}

\subsection{Hyperparameter resilience} \label{sec:hyperparameter_resilience}

Hyperparameter resilience is important in ensuring robust and reliable optimization outcomes, especially in quantum chemistry, where the objective is to find a particular quantum state. A hyperparameter-resilient optimizer increases the chances of successfully finding the optimal solution and reduces the additional overhead of optimizing hyperparameters. 

\begin{figure*}
\centering
\includegraphics{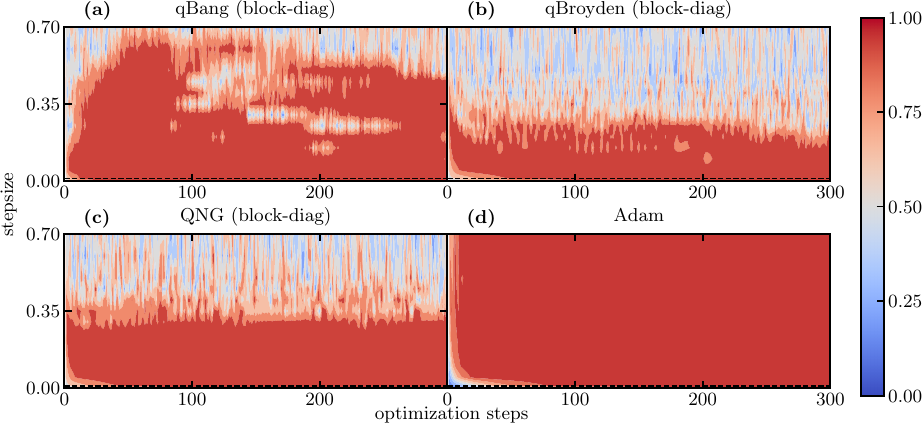}
\caption{Dependence of convergence behavior on the learning rate by comparing the effects of different step sizes on the optimization process. Four optimization algorithms, including \bang{}, \qbroyden{}, QNG with block-diagonal approximation, and Adam, are evaluated with step sizes ranging from 0.01 to 0.7. The optimization performance is assessed using the approximation ratio, which equals one if the energy minimum is reached (see Equation \eqref{eq:approximation_ratio}). A dotted line at a step size of 0.01 is included to facilitate comparison with other simulations.}
\label{fig:vary_stepsize}
\end{figure*}

In \figref{fig:vary_stepsize}, we investigate the effect of varying step size on the approximation ratio over the number of optimization steps in the BP circuit with $9$ qubits and $5$ layers. We use \bang{}, \qbroyden{}, QNG with block diagonal, and Adam as the optimization algorithms and optimize each circuit for $300$ optimization steps. The approximation ratio, equal to one if the energy minimum is reached [see  \eqref{eq:approximation_ratio}], is used to evaluate the optimization performance. We show the approximation ratio plotted against the number of optimization steps for step sizes ranging from $0.01$ to $0.7$. 

\figref{fig:vary_stepsize} demonstrates the greatest strength of Adam -- its extreme resilience. Even for large step-sizes, such as 0.7, Adam remains stable and provides reliable predictions. 
Approximate or perturbative second order optimization methods, such as QNG and \qbroyden{}, are prone to instabilities when using large steps. They tend to result in unreliable predictions for the local curvature which might even further amplify a large step, resulting in oscillating or divergent behaviour. Let us emphasize here that this is not a failure of second-order informed optimization but rather its approximation. Consider for example the step-reducing influence of second-order information in Newtons method for a steep harmonic potential.

Importantly, \bang{} can benefit from the momentum update that it inherits from Adam and achieves a resilience located between Adam and QNG/\qbroyden{}. An even stronger resilience of \bang{} could be realized by unifying the gradient update with the metric update or the use of a more controlled step size depending on the local gradient and cost function, based for example on the Wolfé conditions~\cite{nocedal1999numerical}. 
Given the excellent performance in the previous section, we conclude that \bang{} is a promising optimizer that strikes the balance between low cost, high stability, speed, and accuracy.

\subsection{Noise resilience}
\label{sec:shot_noise}
Understanding the resilience of quantum algorithms to various types of noise is crucial in the noisy intermediate-scale quantum (NISQ) era. Shot noise is one of the most fundamental contributors and arises due to the statistical nature of quantum measurements. Let us put our previous discussions in this context by considering first a simple BP circuit with $9$ qubits and $6$ layers, similar to the setup in Sec.~\ref{sec:barren_plateau_circuit}. The step size is fixed at $\eta=0.01$, and the results are averaged over $15$ random initializations of parameters with 500 shots for each circuit evaluation. 

\begin{figure}
	\centering
	\includegraphics{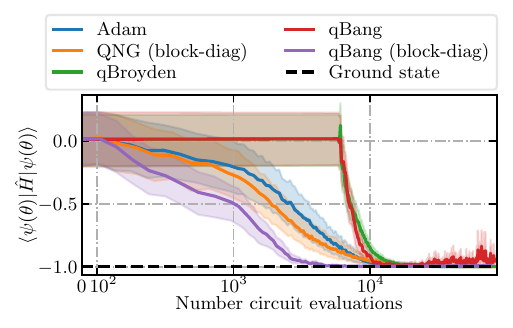}
	\caption{Comparison of optimization performance of Adam, QNG, \qbroyden{}, and \bang{} in finding the ground state of the BP circuit under the influence of shot noise. $\langle \psi(\boldsymbol{\theta}) | \hat H | \psi(\boldsymbol{\theta}) \rangle$ is shown as a function of the number of circuit evaluations. The step size is fixed at $\eta=0.01$, and the results are averaged over $15$ random initializations of parameters. The PQC used consists of $6$ layers. For each evaluation 500 shots are used. 
		The initial plateau in the optimization using \qbroyden{} and \bang{} arises from the significant cost of initially measuring the QFIM.
	}
	\label{fig:shot_noise_barren}
\end{figure}

\autoref{fig:shot_noise_barren} demonstrates that all optimizers exhibit performance closely resembling that of exact state vector simulations. Among them, \bang{} consistently finds the solution most efficiently. 
We note that with shot noise, the estimate of the initial  QFIM is not guaranteed to be positive semi-definite. If necessary, we ensure invertibility (and thus positive definiteness) of the initial QFIM by shifting the diagonal by the most negative eigenvalue $\lambda_{\rm{min}}<0$, as $\textbf{F}_{\text{PD}} = \textbf{F} + \left(\gamma_{\rm{reg}}- \lambda_{\text{min}}\right) \mathbb{1}$, see SI. \secref{sec:PSD_QFIM} for details. Here, $\gamma_{\rm{reg}}>0$ is a small regularising parameter to ensure that $\textbf{F}_{\text{PD}}\succ0$.

\begin{figure}
	\centering
	\includegraphics{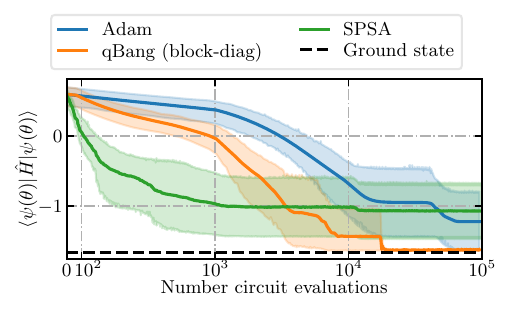}
	\caption{Comparison of optimization performance of SPSA, Adam, and \bang{} with the block-diagonal approximation in finding the ground state of $\rm{H}_4$ using a 2-layer PQC. The expectation value, $\langle \psi(\boldsymbol{\theta}) | \hat H | \psi(\boldsymbol{\theta}) \rangle$ is shown as a function of the  number of circuit evaluations. The step size is fixed at $\eta=0.01$, and 500 shots are used for each evaluation. The results are averaged over $5$ random initializations. Individual trajectories are presented in SI \secref{sec:detailed_h4_shot_noise}.}
	\label{fig:shot_noise_h4}
\end{figure}

Next, we revisit quantum chemistry in the form of the H$_4$ circuit featuring 2 layers, discussed in Sec.~\ref{sec:h4}. 
Circuit evaluations are performed using 500 shots and the results are averaged over 5 random initializations.  We add the Simultaneous Perturbation Stochastic Approximation (SPSA)~\cite{Gacon2021Simultaneous} optimizer, often used in a noisy circuit setting, to our comparison. All optimizers are run for 700 steps, with the exception of SPSA, which is run for 50000 steps. The step size is fixed at $\eta=0.01$.
Figure~\ref{fig:shot_noise_h4} illustrates how \bang{} outperforms Adam, while SPSA is failing to find the minimum. Surprisingly, the performance of \bang{} is even better when affected by noise, likely due to a slightly larger effective step when PD is enforced. Individual trajectories are presented in SI \secref{sec:detailed_h4_shot_noise}.
We can expect the improved performance of \bang{} to be thus of practical relevance for NISQ devices. 

SPSA is a representative of a stochastic approach to optimization, closely related to random walk algorithms, and we refer the reader to Ref.~\cite{Gacon2021Simultaneous,Kubler2020AdaptiveOptimizerMeasurementFrugal} for a detailed discussion and possible improvements. The isolated example shown here is of anecdotal evidence and does not allow to draw any conclusion about the superiority of stochastic or gradient-based approaches. We are indeed convinced that a synergistic approach could be the most promising.

\section{Conclusion}\label{sec:conclusion}

Quantum computing has developed into a vibrant research domain, promising nothing less than a revolution. If this ambitious target can be met depends largely on the availability of fault-tolerant hardware and efficient algorithmic design.
VQAs, merging quantum evaluations on short circuits with classical optimization of the parameterized state, are a promising framework for the use of near-term quantum computing resources. However, associated energy landscapes often feature  sizeable flat areas that are challenging to maneuver. 
Here, we have introduced \bang{} and \qbroyden{}, curvature-informed gradient-based algorithms that perform better than previous approaches for relevant quantum circuits while requiring comparably few evaluations on the QPU. The reduction in quantum evaluations is achieved by performing rank-1 updates to the Fisher information matrix.
Additionally, \bang{} utilizes a momentum-based update rule, providing an additional boost in performance and resilience to changes in hyperparameters. 
We provide access to \bang{} and \qbroyden{} via the freely accessible repository~\cite{fitzek2023qbang}. 

Our benchmarks, including QNG and Adam, are evaluated on a broad range of VQAs. First, we demonstrated for a set of BP circuits~\cite{McClean2018Barrenplateausquantum} that \bang{} is able to tackle flat energy landscapes efficiently. Second, we investigate classical optimization on QAOA circuits in the form of the max-cut problem, resulting in an overall underwhelming performance of all optimizers.
Third, we moved on to quantum chemistry, arguably the most promising application for quantum computing. The associated VQEs have been investigated for three chemical compounds, namely $\rm{H}_4$, LiH, and $\rm{H}_2\rm{O}$, where \bang{} is consistently more  efficient than its competitors.
Lastly, we illustrate that \bang{}, i.e., the combination of \qbroyden{} and Adam, does indeed lead to a more noise- and hyper-parameter-resilient optimizer than QNG or \qbroyden{} itself.
\bang{} is an efficient and capable optimizer, yet the strongest aspect of our work is that it inspires a new generation of optimizers -- \bang{} representing a first step in an evolutionary process.  Such an evolution will be fostered by understanding the consequences of locality, complexity, and entanglement on the existence of BPs~\cite{ragone2023unified, Fontana2023}.

With the increasing number of qubits and their connectivity, the number of quantum Ansatz parameters will grow, resulting in increasing pressure on the classical optimizers.
With this in mind, we suggest using \bang{} as a ``convergence starter'' for optimization problems that involve a sizeable number of Ansatz layers. One potential approach is to optimize the first few layers and then keep those optimized layers with their parameters as an initial guess for the next few layers to optimize. This process can be repeated recursively until all layers are optimized and could significantly reducing the number of optimization steps required to find an acceptable ground-state energy. For a last refinement, one could use the VarQITE algorithm or restart the \bang{} algorithm by wiping the memory. 
Furthermore, the Fisher information matrix encodes information about the degree of linear dependence, i.e., it can be used to maximize the efficiency of additional layers and improve stability by controlling over-parametrization~\cite{Larocca_2023}.
To this end, it should be noted that an application to relevant problems with real-world devices remains a considerable challenge.

\vspace{1em}
\begin{acknowledgments}
We thank Anton Frisk Kockum, Mats Granath, Leo Laine, Davide Castaldo, and G\"oran Johansson for insightful discussions.
This work was supported by the Swedish Research Council (VR) through Grant No. 2016-06059 and the computational resources provided by
the Swedish National Infrastructure for Computing at Chalmers Centre for Computational Science and Engineering partially funded by the Swedish Research Council through grant agreement no. 2018-05973. D.F. and R.S.J. acknowledge the Knut and Alice Wallenberg (KAW) Foundation for funding through the Wallenberg Centre for Quantum Technology
(WACQT). 
W.D. and C.S. acknowledge funding from the Horizon Europe research and
innovation program of the European Union under the Marie
Sk{\l}odowska-Curie grant agreement no.\ 101062864 and 101065117. 
Partially funded by the European Union. 
Views and opinions expressed are, however, those of the author(s) only and do not necessarily reflect those of the European Union or REA. 
Neither the European Union nor the granting authority can be held responsible for them.

\end{acknowledgments}

\bibliographystyle{quantum}
\bibliography{library}

\onecolumn\newpage
\appendix

\textbf{\huge Appendix} 
\section{Single trajectories including QITE} \label{app:single_trajectories}

\begin{figure}[!h]
    \centering
    \includegraphics{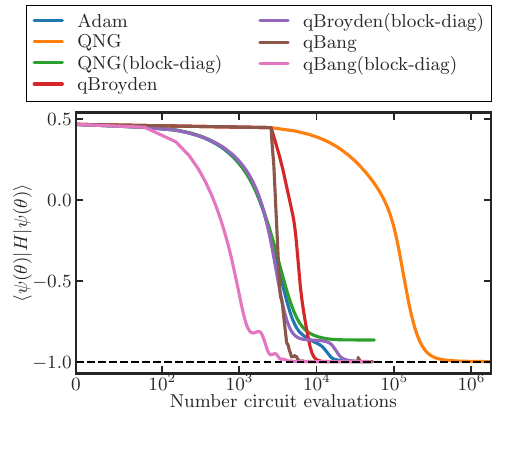}
    \caption{Comparison of optimization performance of Adam, QNG, \qbroyden{}, and \bang{} in finding the ground state of the BP circuit. $\langle \psi(\boldsymbol{\theta}) | \hat H | \psi(\boldsymbol{\theta}) \rangle$ is shown as a function of the  number of circuit evaluations. The step size is fixed at $\eta=0.01$. The PQCs used consist of $4$ layers. All optimizers perform 700 steps, which results in a wide range of circuit evaluations due to the expensive estimation of the Fisher information. The initial plateau in the optimization using QNG, \qbroyden{} and \bang{} arises from the significant cost of initially measuring the QFIM.
    }
    \label{fig:comparing_optimizers_with_full_varqite}
\end{figure}

In this section, we compare the performance of \bang{}, \qbroyden{}, QNG, and Adam optimizers, including QNG using the full quantum Fisher information matrix (QFIM) at each step. We consider a barren plateau (BP) circuit with 4 layers and 9 qubits, resulting in 36 tunable parameters. We optimize for 700 steps, resulting in varying circuit evaluations since the QFIM requires $n_\theta^2$ circuit evaluations while approximations such as diagonal or block-diagonal approximation require only $n_\theta + l$ circuit evaluations, where $n_\theta$ is the number of variational parameters and $l$ is the number of layers in the circuit. QNG using the QFIM is equivalent, up to a constant factor, to VarQITE~\cite{McArdle2019Variationalansatzbasedquantum}. QNG, qBang, and qBroyden require the QFIM in the first step, explaining the initial plateau in the number of circuit evaluations compared to Adam or the approximated versions. All optimizers, except for QNG with the block-diagonal approximation, converge to the exact ground state solution. The results in \figref{fig:comparing_optimizers_with_full_varqite} show that a single estimate of the QFIM, in combination with an appropriate cost-efficient metric update, is sufficient to speed up convergence to the desired ground state. 

\subsection{Why updating the metric is important (ablation study)} \label{sec:update_rule_ablation_study}

In this subsection, we perform an ablation study to investigate the impact of the update rule formula on optimization performance. We use a BP circuit with $9$ qubits and $6$ layers and average over $10$ random parameter initializations. 

\begin{figure}
    \centering
    \includegraphics{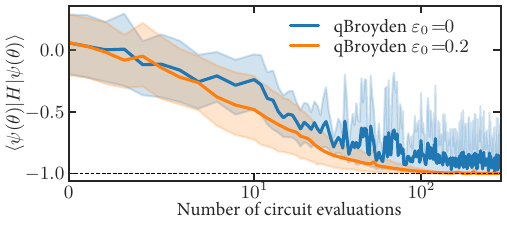}
    \caption{Effect of metric update on the optimization performance in a 6-layer, 9-qubit BP circuit. The performance of \qbroyden{} is compared for $\varepsilon_0=0$ and $\varepsilon_0=0.2$. When $\varepsilon_0=0$, the update rule~\eqref{eq:updaterule} is not used. For both settings, the algorithms are initialized with the full Fisher information matrix. Results are averaged over 10 random parameter initializations with 300 optimization steps each.}
    \label{fig:ablation}
\end{figure}

We show in \figref{fig:ablation} that, for the first iterations, both algorithms perform similarly, but in the long run, without a metric update, oscillations appear in the system, leading to no convergence of the optimization.
To understand this behavior, let us recall that the Fisher information is a measure of how much a parametrized state changes under a change of a parameter~\cite{Meyer2021}. This information \textcolor{black}{can be understood as} an adaptive step size for each parameter to optimize. However, since the energy landscape changes during optimization, we need to adjust the Fisher information to ensure proper convergence. As shown in Figure~\ref{fig:ablation}, if we do not correct the metric, oscillations start after a few optimization steps when the energy landscape has undergone a sufficient change and is no longer described by the initial QFIM. On the other hand, the quasi-Newton updates to the initial metric ensure that the gradient descent is more consistent and \qbroyden{} find the ground state quickly.

The update rule is thus crucial and provides the necessary correction to adjust the curvature of the Fisher information matrix based on the current point in the energy landscape. This has two significant advantages. First, it reduces the number of circuit queries required, and second, it simplifies the algorithm's execution on the hardware because we only need to estimate the Fisher information once on the quantum device.

In summary, the ablation study in Fig.~\ref{fig:ablation} shows that correcting the metric is essential to avoid oscillations and ensure convergence of the optimization process.

{\color{black}
\subsection{Analysis of H$_4$ optimization trajectories under shot noise}
\label{sec:detailed_h4_shot_noise}
Revisiting the H$_4$ circuit with 2 layers, as discussed in \secref{sec:shot_noise} of the main document, we now shift our focus from averaged results to an examination of individual optimization trajectories. This approach provides a more granular view of the optimizer performance under shot noise conditions. Each circuit evaluation is performed using 500 shots, and we observe the behavior across 5 random initializations.

\begin{figure}[h]
    \centering
    \includegraphics{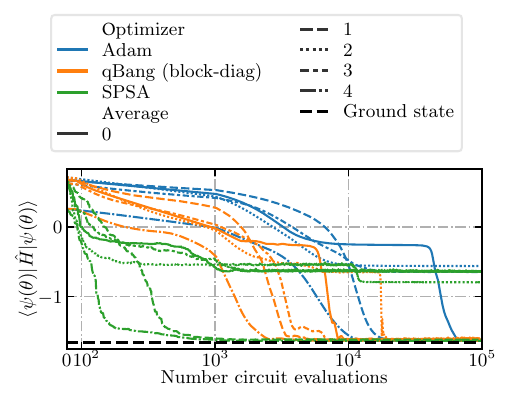}
    \caption{Individual optimization trajectories for the H$_4$ circuit with a 2-layer PQC. The expectation value, $\langle \psi(\boldsymbol{\theta}) | \hat H | \psi(\boldsymbol{\theta}) \rangle$, is shown for each circuit evaluation. The step size is fixed at $\eta=0.01$. Each line represents a separate optimization run, illustrating the variability among trajectories.}
    \label{fig:individual_shot_noise_h4}
\end{figure}

In the analysis, represented in Fig.~\ref{fig:individual_shot_noise_h4}, \bang{} demonstrates reliable performance in finding the ground state and outperforms the Adam and Simultaneous Perturbation Stochastic Approximation (SPSA) optimizer. Notably, SPSA, despite running for 50000 steps, struggles to locate the minimum in several cases.
}

\section{Circuit layouts and Hamiltonians}

This section collects all the circuit ansätze and Hamiltonian descriptions used for the benchmarks. All of the circuits are built with $l$ layers. The more layers the larger the expressivity of the circuit which allows for potentially more accurate solutions but also increases the linear dependence of parameters. All circuits are optimized in a closed-loop with a classical optimization algorithm to minimize $\bra{\psi(\boldsymbol{\theta})} \hat{H}\ket{\psi(\boldsymbol{\theta})}$, where $\psi(\boldsymbol{\theta})$ describes the circuit ansatz. 

\subsection{Barren plateau circuit} 

\begin{figure}
    \centering
    \includegraphics{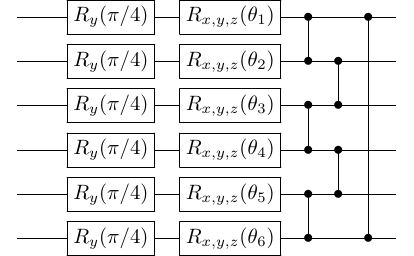}
    \caption{The BP circuit ansatz. The ansatz consists of an initial layer of $R_y(\pi/4)$ gates followed by $l$ layers of parameterized Pauli rotations and a controlled-$Z$ entangling layer, initialized in the state $\ket{0}^n$ for all $n$ qubit registers.}
    \label{fig:barren_plateau_prone_circuit_ansatz}
\end{figure}

BPs are a major obstacle in quantum computing, hindering its potential for solving complex problems~\cite{McClean2016, Cerezo2021Costfunctiondependent}. 
The BP circuit is an example of this phenomenon and utilizes the objective operator $\hat H=\hat Z_1\hat Z_2$ with a ground state energy of $-1$. The circuit is initialized in the state $\ket{0}^n$ and consists of an initial fixed layer of $R_y(\pi/4)$ gates acting on $n$ qubits, followed by $l$ layers of parameterized Pauli rotations with an entangling layer of controlled-$Z$ gates, as shown in Figure \ref{fig:barren_plateau_prone_circuit_ansatz}.
This circuit is a critical benchmark for understanding and addressing the BP problem in quantum computing.

\subsection{Quantum approximate optimization algorithm circuit ansatz}

The Quantum Approximate Optimization Algorithm (QAOA) is a quantum algorithm that can be used to solve combinatorial optimization problems. One such problem is the max-cut problem, which involves partitioning a set of vertices in a graph into two disjoint subsets such that the number of edges between the subsets is maximized~\cite{Farhi2014QuantumApproximateOptimization}.

The max-cut problem is mapped onto a quantum optimization problem by constructing a cost Hamiltonian $\hat{H}_C$ that encodes the objective function of the max-cut problem. The cost Hamiltonian is defined as follows:
\begin{equation}
    \hat{H}_C = \sum_{(i,j)\in E} \frac{1}{2}(\hat{\mathds{1}}- \hat{Z}_i \hat{Z}_j),
\end{equation}
where $E$ is the set of edges in the graph, and $Z_i$ and $Z_j$ are the Pauli Z operators acting on the qubits corresponding to vertices $i$ and $j$, respectively. The cost Hamiltonian penalizes states in which neighboring vertices are in the same subsets since the corresponding edge contributes $1$ to the energy in these states.

The quantum circuit uses two non-commuting operators, the cost Hamiltonian and the mixing Hamiltonian, to evolve the system towards states that optimize the cost function. The mixing Hamiltonian is typically a sum of Pauli $X$ operators, acting as a \enquote{driver} that moves the system away from the initial state and encourages exploration of different states.

\begin{figure}
    \centering
    \includegraphics{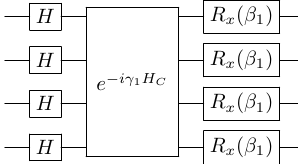}
    \caption{The QAOA circuit ansatz. It is composed of alternating layers of the cost Hamiltonian and the mixing Hamiltonian. The circuit is initialized in the state $\ket{0}^n$, where $n$ is the number of qubits required by the cost Hamiltonian. The parameters of the circuit are optimized to maximize the expected value of the cost function.}
    \label{fig:qaoa_circuit}
\end{figure}

Figure \ref{fig:qaoa_circuit} shows a QAOA circuit ansatz with one layer, applying the cost and mixing Hamiltonians. The circuit is initialized in the state $\ket{0}^n$, which is transformed into the uniform superposition state $\ket{+}^n$ via the Hadamard gate. The QAOA provides an approximation to the optimal solution, with the quality of the approximation expected to improve as the number of layers $l$ is increased.

\subsection{Chemistry applications} \label{app:chemistry_applications}

\begin{figure}
    \centering
    \includegraphics{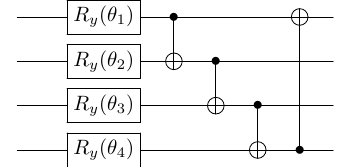}
    \caption{The hardware efficient circuit ansatz is composed of $l$ layers of parametrized single qubit $R_y$ rotations and a ring of CNOT gates to entangle the qubits. The circuit is applied to $n$ qubits, with the parameters optimized to minimize the energy of the molecular system.}
\label{fig:hardware_efficient_ansatz_simple}
\end{figure}

We employed minimal basis sets (STO-6G) for all quantum chemistry problems and used a frozen core approximation for LiH and H$_2$O (with the 1s orbital of Li and O, respectively, frozen)~\cite{Helgaker2000}. 
We used a hardware-efficient ansatz (HEA)~\cite{Kandala2017HardwareefficientVariationalQuantum} that utilizes $8$, $10$, and $12$ qubits for $\rm{H}_4$, LiH, and $\rm{H}_2\rm{O}$, respectively. This ansatz is composed of $l$ layers, each comprising a tunable $R_y(\theta)$ gate on each qubit register, followed by a closed ring of CNOT gates. A 1-layer motif of the HEA for 4 qubits can be seen in Fig.~\ref{fig:hardware_efficient_ansatz_simple}. 
In the following, we list the geometries of all the studied molecular problems (in the \texttt{xyz}-format and atomic units):
\begin{lstlisting}[caption={H$_4$ geometry in \texttt{xyz}-format and atomic units}]
4
*
H     2.1213	 2.1213     0.0
H     2.1213	-2.1213     0.0
H    -2.1213	 2.1213     0.0
H    -2.1213	-2.1213     0.0
\end{lstlisting}

\begin{lstlisting}[caption={LiH geometry in \texttt{xyz}-format and atomic units}]
2
*
Li     0.0   0.0   0.0
H      0.0   0.0   3.0
\end{lstlisting}

\begin{lstlisting}[caption={H$_2$O geometry in \texttt{xyz}-format and atomic units}]
3
*
O   0.0        0.0        0.0
H   0.8081     1.0437     0.0
H   0.8081    -1.0437     0.0
\end{lstlisting}

We provide a python implementation of the circuits and Hamiltonians used in this work in~\cite{fitzek2023qflow}.

\subsubsection{Hardware-efficient $R_y$ Ansatz}
HEAs, like the $R_y$ Ansatz shown in Fig.~\ref{fig:hardware_efficient_ansatz_simple}, are commonly used in quantum computing studies of chemical and physical systems. 
It is, however, not trivial and thus an active field of research 
how increasing the number of layers affects the \enquote{expressivity} -- how well $\ket{\psi(\boldsymbol{\theta})}$ can approximate the target $\ket{\Psi}$ -- of a HEA~\cite{Sim2019, Du2020, Funcke2021, Du2022, DCunha2023}.
This effect can be seen in the slow convergence of the total energy of H$_4$ with the number of ansatz layers, see Fig.~\ref{fig:hardware_efficient_ansatz_simple}.
Nevertheless, we chose to study HEA in this work since (a) they are desirable to use as they lead to smaller errors due to hardware noise~\cite{Kandala2017HardwareefficientVariationalQuantum}.
However, especially because it was proven that the gradient exponentially vanishes for deep, randomly initialized HEA~\cite{McClean2018Barrenplateausquantum, Holmes2022}.

\section{Initialization using Hartree-Fock parameters} \label{app:initialization_using_hartree_fock}

\begin{figure*}
    \centering
    \includegraphics{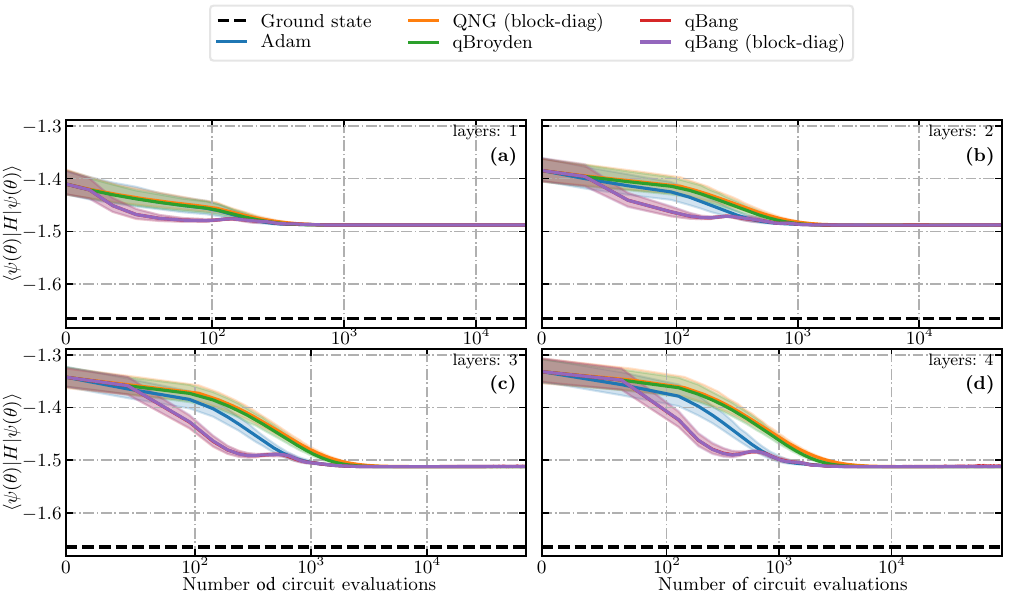}
    \caption{Comparing the optimization performance for the four optimizers, Adam, Quantum natural gradient (QNG) with block-diag approximation, \qbroyden{} with full QFIM and \bang{} with block-diag and full QFIM for finding the ground state of $\rm{H}_4$ with a variational quantum circuit. The step size is set to 0.01. We average over 15 randomly perturbed HF-parameter initializations. We use a PQC with 1, 2, 3, and 4 layers.}
    \label{fig:h4_hf}
\end{figure*}

We present the performance starting from the Hartree-Fock parameter initialization in \figref{fig:h4_hf}. We compare the optimization performance of four different optimizers, namely Adam, Quantum natural gradient (QNG) with block-diag approximation, \qbroyden{} with full QFIM and \bang{} with block-diag and full QFIM, for finding the ground state of $\rm{H}_4$ using a variational quantum circuit. The step size for each optimizer is set to 0.01. We employ a parameterized quantum circuit (PQC) with varying numbers of layers, from 1 to 4, to explore the impact of circuit depth on the optimization performance of each optimizer. To ensure the robustness of our results, we perform 15 independent optimization runs, each with a randomly perturbed Hartree-Fock parameter initialization. Overall we see stable convergence behavior for the chosen circuit ansatz. 
All optimizers converge to the same minimum.

\section{Collection of Algorithms}

This section summarizes all the optimization algorithms introduced in this work. 

\qbroyden{} is a quasi-Newton method that approximates the  QFIM matrix using rank-one updates. In each iteration, the inverse QFIM is updated using an updating rule that depends on the gradient and parameter differences between the current and previous iterations. Algorithm~\ref{alg:qbroyden} presents the pseudo-code for \qbroyden{}.

\bang{} is an extension of \qbroyden{} that incorporates both the approximation of the QFIM and momentum. In each iteration, the gradients are first normalized using the adaptive moment estimation (Adam) method, and then a preconditioned gradient step is taken using the inverse QFIM. Similar to \qbroyden{}, \bang{} can also incorporate QNG, QFIM, or the identity matrix as a preconditioner. Algorithm~\ref{alg:bang} presents the pseudo-code for \bang{}.

Momentum QNG combines momentum optimization with QNG. In each iteration, we utilize an Adam~\cite{Kingma2017AdamMethodStochastic} inspired update for the momentum and then take a natural gradient step by using both the momentum and the QNG approximation of the QFIM. Algorithm~\ref{alg:momentumqng} presents the pseudo-code for Momentum QNG. 

A Python implementation for all three optimization algorithms can be found in~\cite{fitzek2023qbang}. 

\begin{algorithm}[H]
    \caption{\qbroyden{} }
    \begin{algorithmic}[1]
    \State  \textbf{Input:} learning rates $\eta=0.01$, $\varepsilon_0=0.2$
    \State  \textbf{Input:} convergence criterion $\gamma$
    \State  \textbf{Input:} PQC $U(\boldsymbol{\theta})$ with initial parameter vector $\boldsymbol{\theta}_0\in \mathbb{R}^p$.
    \State \textbf{Initialization:}   $k\leftarrow0$, $\mathbf{B}_{0}^{-1}$ via QNG, QFIM or Identity 
    \State \texttt{not\_converged}$\leftarrow$\texttt{true} 
    \While{\texttt{not\_converged}}     
        \State \textbf{QC}: estimate $\mathcal{L}(\boldsymbol{\theta}_k)$
        \State \textbf{QC}: estimate $\nabla \mathcal{L}(\boldsymbol{\theta}_k)$
        \If{$\|\mathbf{B}_k^{-1} \nabla \mathcal{L}(\boldsymbol{\theta}_k)\|_2>\gamma$}
            \State $\boldsymbol{\theta}_{k+1} \leftarrow \boldsymbol{\theta}_{k} - \eta \, \mathbf{B}_k^{-1} \nabla \mathcal{L}(\boldsymbol{\theta}_k) $ 
            \State $\varepsilon_k\leftarrow \frac{\varepsilon_0}{k+1}$        
            \State $\mathbf{B}_{k+1}^{-1}\leftarrow$  \eqref{eq:updaterule}
            \State $k\leftarrow k+1$
        \Else     
            \State \texttt{not\_converged}$\leftarrow $\texttt{false} 
            \State      $\boldsymbol{\theta}^*\leftarrow\underset{\{\boldsymbol{\theta}_n\}_0^k }{\operatorname{argmin}} \,\mathcal{L}(\boldsymbol{\theta}_n)$              
        \EndIf
    \EndWhile
    \State{\Return $\boldsymbol{\theta}^*$}
    \end{algorithmic}
    \label{alg:qbroyden2}
\end{algorithm}

\begin{algorithm}[H]
    \caption{\bang{} }
    \begin{algorithmic}[1]
    \State  \textbf{Input:} learning rates $\eta=0.01$, $\varepsilon_0=0.2$
    \State  \textbf{Input:} decay rates $\beta_1=0.9$ and $\beta_2=0.999$
    \State  \textbf{Input:} convergence criterion $\gamma$
    \State  \textbf{Input:} PQC $U(\boldsymbol{\theta})$ with initial parameter vector $\boldsymbol{\theta}_0\in \mathbb{R}^p$.
    \State \textbf{Initialization:}   $k\leftarrow0$, $\boldsymbol{m}_{-1} \leftarrow \mathbf{0}$, $\boldsymbol{v}_{-1} \leftarrow \mathbf{0}$, $\mathbf{B}_{0}^{-1}$ via QNG, QFIM or Identity 
    \State \texttt{not\_converged}$\leftarrow$\texttt{true} 
    \While{\texttt{not\_converged}}     
        \State \textbf{QC}: estimate $\mathcal{L}(\boldsymbol{\theta}_k)$
        \State \textbf{QC}: estimate $\nabla \mathcal{L}(\boldsymbol{\theta}_k)$
        \State $\boldsymbol{m}_k\leftarrow\beta_{1} \boldsymbol{m}_{k-1}+\left(1-\beta_{1}\right) \nabla \mathcal{L}(\boldsymbol{\theta}_k)$ 
        \State $\boldsymbol{v}_k \leftarrow \beta_{2}  \boldsymbol{v}_{k-1}+\left(1-\beta_{2}\right)  \nabla \mathcal{L}(\boldsymbol{\theta}_k) \odot \nabla \mathcal{L}(\boldsymbol{\theta}_k)$ 
        \State $\boldsymbol{\widehat{m}}_k \leftarrow \boldsymbol{m}_k /\left(1-\beta_{1}^{k+1}\right)$
        \State $\boldsymbol{\widehat{v}}_k \leftarrow \boldsymbol{v}_k /\left(1-\beta_{2}^{k+1}\right)$
        \State $\{\boldsymbol{p}_k\}_l\leftarrow \{\boldsymbol{\widehat{m}}_k \}_l/\left(\sqrt{\{\boldsymbol{\widehat{v}}_k \}_l}+\kappa\right),\  \forall l\in \{1,2,\ldots, p\}$ 
        \If{$\|\mathbf{B}_k^{-1}\boldsymbol{p}_k\|_2>\gamma$}
            \State $\boldsymbol{\theta}_{k+1} \leftarrow \boldsymbol{\theta}_{k} - \eta \, \mathbf{B}_k^{-1}\boldsymbol{p}_k$ 
            \State $\varepsilon_k\leftarrow \frac{\varepsilon_0}{k+1}$        
            \State $\mathbf{B}_{k+1}^{-1}\leftarrow$  \eqref{eq:updaterule}
            \State $k\leftarrow k+1$
        \Else     
            \State \texttt{not\_converged}$\leftarrow $\texttt{false} 
            \State      $\boldsymbol{\theta}^*\leftarrow\underset{\{\boldsymbol{\theta}_n\}_0^k }{\operatorname{argmin}} \,\mathcal{L}(\boldsymbol{\theta}_n)$              
        \EndIf
    \EndWhile
    \State{\Return $\boldsymbol{\theta}^*$}
    \end{algorithmic}
    \label{alg:bang2}
\end{algorithm}

\begin{algorithm}[H]
    \caption{Momentum QNG}
    \begin{algorithmic}[1]
    \State  \textbf{Input:} learning rates $\eta=0.01$, $\varepsilon_0=0.2$
    \State  \textbf{Input:} decay rates $\beta_1=0.9$ and $\beta_2=0.999$
    \State  \textbf{Input:} convergence criterion $\gamma$
    \State  \textbf{Input:} PQC $U(\boldsymbol{\theta})$ with initial parameter vector $\boldsymbol{\theta}_0\in \mathbb{R}^p$.
    \State \textbf{Initialization:}   $k\leftarrow0$, $\boldsymbol{m}_{-1} \leftarrow \mathbf{0}$, $\boldsymbol{v}_{-1} \leftarrow \mathbf{0}$ 
    \State \texttt{not\_converged}$\leftarrow$\texttt{true} 
    \While{\texttt{not\_converged}}     
        \State \textbf{QC}: estimate $\mathcal{L}(\boldsymbol{\theta}_k)$
        \State \textbf{QC}: estimate $\nabla \mathcal{L}(\boldsymbol{\theta}_k)$
        \State \textbf{QC}: estimate $\mathbf{B}_{k}$
        \State $\boldsymbol{m}_k\leftarrow\beta_{1} \boldsymbol{m}_{k-1}+\left(1-\beta_{1}\right) \nabla \mathcal{L}(\boldsymbol{\theta}_k)$ 
        \State $\boldsymbol{v}_k \leftarrow \beta_{2}  \boldsymbol{v}_{k-1}+\left(1-\beta_{2}\right)  \nabla \mathcal{L}(\boldsymbol{\theta}_k) \odot \nabla \mathcal{L}(\boldsymbol{\theta}_k)$ 
        \State $\boldsymbol{\widehat{m}}_k \leftarrow \boldsymbol{m}_k /\left(1-\beta_{1}^{k+1}\right)$
        \State $\boldsymbol{\widehat{v}}_k \leftarrow \boldsymbol{v}_k /\left(1-\beta_{2}^{k+1}\right)$
        \State $\{\boldsymbol{p}_k\}_l\leftarrow \{\boldsymbol{\widehat{m}}_k \}_l/\left(\sqrt{\{\boldsymbol{\widehat{v}}_k \}_l}+\kappa\right),\  \forall l\in \{1,2,\ldots, p\}$ 
        \If{$\|\mathbf{B}_k^{-1}\boldsymbol{p}_k\|_2>\gamma$}
            \State $\boldsymbol{\theta}_{k+1} \leftarrow \boldsymbol{\theta}_{k} - \eta \, \mathbf{B}_k^{-1}\boldsymbol{p}_k $
            \State $k\leftarrow k+1$
        \Else     
            \State \texttt{not\_converged}$\leftarrow $\texttt{false} 
            \State      $\boldsymbol{\theta}^*\leftarrow\underset{\{\boldsymbol{\theta}_n\}_0^k }{\operatorname{argmin}} \,\mathcal{L}(\boldsymbol{\theta}_n)$              
        \EndIf
    \EndWhile
    \State{\Return $\boldsymbol{\theta}^*$}
    \end{algorithmic}
    \label{alg:momentumqng}
\end{algorithm}

\section{Relation between Fisher information and Hessian}\label{app:connectionFIMHessian}

For certain classes of classical optimization problems, the natural gradient method is equivalent to the Newton method. Here, we describe a class of problems where the Fisher information matrix (FIM) and Hessian are related. \textcolor{black}{This relationship is well known in the literature, see, e.g., Ref.~\cite{shima2007geometry}.}

Let the random variable $X\in \mathcal{D}_X$ be distributed according to the probability density function $p(X;\boldsymbol{\theta})$, where the distribution is parametrized by the continuous parameter vector $\boldsymbol{\theta}$.  Through the Cramér-Rao lower bound, the FIM describes how well $\boldsymbol{\theta}$ can be estimated, ideally, from observations of $X$.  The FIM is defined as
\begin{equation}
    \mathbf{I}_{i,j} = {\rm E}_X\left[\left(\partial_{\theta_i} \ln p(X;\boldsymbol{\theta})\right)\left(\partial_{\theta_j} \ln p(X;\boldsymbol{\theta})\right)\right],
\end{equation}
or $\mathbf{I}_{i,j} = \int_{\mathcal{D}_X}{\rm d}X\,\frac{(\partial_{\theta_i}p(X;\boldsymbol{\theta}))(\partial_{\theta_j}p(X;\boldsymbol{\theta}))}{p(X;\boldsymbol{\theta})} $.
A required   condition of regularity permits us to exchange the order of integration and differentiation\footnote{Specifically, the required condition is that $\int_{\mathcal{D}_x}{\rm d}X \partial_{\theta_i}\partial_{\theta_j} p(X;\boldsymbol{\theta})= 0$,   which is satisfied if $\mathcal{D}_X$ is independent of $\boldsymbol{\theta}$.} and the FIM can then be described with the second order derivatives, as $
        \mathbf{I}_{i,j} = -{\rm E}_X\left[  \frac{\partial^2}{\partial_{\theta_i}\partial_{\theta_j}} \ln p(X;\boldsymbol{\theta})\right].$
        
Let us assume we are dealing with a stochastic optimization problem, where the task is to minimize some loss function $L$. That is, we want to minimize the expectation over $X$ of some parametrized error function $l(X;\boldsymbol{\theta})$, as $L={\rm E}_X[l(X;\boldsymbol{\theta})]$.  Newton-based optimization  involves the Hessian of $L$, which has elements $\mathbf{H}_{i,j}=\frac{\partial^2}{\partial_{\theta_i}\partial_{\theta_j}}L$, or 
\begin{align}
    \mathbf{H}_{i,j}&=\frac{\partial^2}{\partial_{\theta_i}\partial_{\theta_j}} {\rm E}_X[l(X;\boldsymbol{\theta})] \\
    &= \int_{\mathcal{D}_X}{\rm d}X\, \frac{\partial^2}{\partial_{\theta_i}\partial_{\theta_j}}l(X;\boldsymbol{\theta})p(X;\boldsymbol{\theta})
\end{align}
because we require that we can exchange order of integration and differentiation. Assume now that the error function can be written as $l(X;\boldsymbol{\theta}) = b(\boldsymbol{\theta})-\ln c(X;\boldsymbol{\theta})$, for some functions $b$ and $c$. If $l(X;\boldsymbol{\theta})\geq 0$ for all $X,\boldsymbol{\theta}$, this implies $0\leq b(\boldsymbol{\theta})$ and $0<c(X;\boldsymbol{\theta})\leq 1$. As a consequence,
\begin{align}
    \mathbf{H}_{i,j}&=\int_{\mathcal{D}_X}{\rm d}X\,\frac{\partial^2}{\partial_{\theta_i}\partial_{\theta_j}}b(\boldsymbol{\theta})p(X;\boldsymbol{\theta}) \\
    &-\int_{\mathcal{D}_X}{\rm d}X\,\frac{\partial^2}{\partial_{\theta_i}\partial_{\theta_j}}\ln c(X;\boldsymbol{\theta})p(X;\boldsymbol{\theta}).
\end{align}
Using $\partial_{\theta_i}\int p(X;\boldsymbol{\theta}))=0$, the Hessian reduces to the FIM in the particular case that
\begin{align}
    c(X;\boldsymbol{\theta})  &=p(X;\boldsymbol{\theta}), \\
    \frac{\partial^2}{\partial_{\theta_i}\partial_{\theta_j}} b(\boldsymbol{\theta}) &= \int_{\mathcal{D}_X}{\rm d}X\, \ln p(X;\boldsymbol{\theta})\frac{\partial^2}{\partial_{\theta_i}\partial_{\theta_j}} p(X;\boldsymbol{\theta}),
 \end{align}
 i.e., the Hessian and Fisher information matrix overlap exactly. In practice, this means that a class of problems where the natural gradient method is equivalent to the Newton method are those where  the probability density function is exponential in the error function, i.e., $p(X;\boldsymbol{\theta})=\exp(b(\boldsymbol{\theta})-l(X;\boldsymbol{\theta})))$. 
 The connection between Fisher information and Hessian has been utilized before in the domain of neural network optimization with Gaussian statistics~\cite{Amari1998NaturalGradientWorks,amari2016information}.

The above relation for the classical Fisher information matrix and Hessian takes for variational quantum algorithms the form $\mathcal{L} = \int {\rm d}X\, p(X;\boldsymbol{\theta}) l(X;\boldsymbol{\theta})$ with probability density function $p(X;\boldsymbol{\theta}) = \Psi^*(X;\boldsymbol{\theta})\Psi(X;\boldsymbol{\theta})$ and energy density $ l(X;\boldsymbol{\theta}) =  \Psi^*(X;\boldsymbol{\theta})\hat{H}\Psi(X;\boldsymbol{\theta}) /p(X;\boldsymbol{\theta})$. 

\section{  Properties of the approximate metric}\label{app:normalized_step}

For the optimization algorithms we have introduced, the update rule
\begin{align}
    \mathbf{B}_{k+1} = (1-\varepsilon_k)\mathbf{B}_{k}+\varepsilon_k\nabla \mathcal{L}_k\nabla \mathcal{L}_k^\top. 
\end{align}
is applied to iterate on the metric. \textcolor{black}{If the initial matrix $\mathbf{B}_0$ is positive semi-definite ($\mathbf{B}_0\succeq0$), the update rule preserves this property for all $\mathbf{B}_k$. To see this, first assume $\mathbf{B}_k\succeq 0$. Then it holds that $(1-\varepsilon_k)\mathbf{B}_k\succeq 0$ for all $\varepsilon_k\in (0,1)$. Next, $\varepsilon_k\nabla\mathcal{L}_k\nabla\mathcal{L}_k^\top\succeq0$ for all $\varepsilon_k>0$, because 
\begin{align}
    \boldsymbol{x}^\top \nabla\mathcal{L}_k \nabla\mathcal{L}_k^\top \boldsymbol{x} &= \langle\boldsymbol{x},\nabla\mathcal{L}_k\rangle\langle\nabla\mathcal{L}_k , \boldsymbol{x}\rangle \\
    &=  \langle\nabla\mathcal{L}_k , \boldsymbol{x}\rangle ^2 \geq0
\end{align}
for all $\boldsymbol{x}$. 
The sum of two matrices that are positive semi-definite is again positive semi-definite. Additionally, if we initialise $\mathbf{B}_0\succ0$, we preserve $\mathbf{B}_{k}\succ0$ for all $k$. Consequently, it follows that $\mathbf{B}_{k}^{-1}$ exists and is positive definite for all $k$.}

To \textcolor{black}{provide further intuition for  the algorithm, we study the long-term behavior of the metric under the update  rule.} Each step taken \textcolor{black}{in the parameter space} is defined by the vector $\Delta_{k} = \mathbf{B}_{k}^{-1}\nabla \mathcal{L}(\boldsymbol{\theta}_k)$. Now, we insert in the $\Delta_{k+1}$ explicitly the expression
\begin{align}
    \mathbf{B}_{k+1}^{-1} 
    =& \left[\mathbb{1} - \frac{\varepsilon_k\mathbf{B}_k^{-1}\nabla \mathcal{L}_k\nabla \mathcal{L}_k^\top}{1-\varepsilon_k(1-\nabla \mathcal{L}_k^\top \mathbf{B}_k^{-1} \nabla \mathcal{L}_k)}\right] \frac{ \mathbf{B}_k^{-1}}{1-\varepsilon_k}\label{eq:updaterule2} 
\end{align}
to get
\begin{align}
\Delta_{k+1} &= \left[\mathbb{1} - \frac{\varepsilon_k\mathbf{B}_k^{-1}\nabla \mathcal{L}(\boldsymbol{\theta}_k)\nabla \mathcal{L}(\boldsymbol{\theta}_k)^\top}{1+\varepsilon_k(\nabla \mathcal{L}(\boldsymbol{\theta}_k)^\top \mathbf{B}_k^{-1} \nabla \mathcal{L}(\boldsymbol{\theta}_k)-1)}\right]  \frac{ \mathbf{B}_k^{-1}\nabla \mathcal{L}(\boldsymbol{\theta}_{k+1})}{1-\varepsilon_k}\\
&=  \left[\mathbf{B}_k^{-1} - \frac{\varepsilon_k\Delta_k\Delta_k^\top}{1+\varepsilon_k(\Delta_k^\top\nabla  \mathcal{L}(\boldsymbol{\theta}_k)-1)}\right] \frac{ \nabla \mathcal{L}(\boldsymbol{\theta}_{k+1})}{1-\varepsilon_k}.\label{eq:app_B2}
\end{align}
Since $\lim_{k\rightarrow \infty}\varepsilon_k = 0$,  for sufficiently large $k$ the effective  step is $\Delta_k \approx  \mathbf{B}_{k-1}^{-1}\nabla \mathcal{L}(\boldsymbol{\theta}_k)$. Let us denote  the second term inside the parenthesis of \eqref{eq:app_B2} by $\gamma_k = \varepsilon_k\Delta_k\Delta_k^\top/(1+\varepsilon_k(\Delta_k^\top \nabla \mathcal{L}(\boldsymbol{\theta}_k)-1))$ and refer to it as the \textit{innovation} at each step. Expanding from the initial point and defining $\varepsilon_{-1}=0$,  the generic step can be written
\begin{equation}
    \Delta_k = \left[\mathbf{B}_0^{-1}-\mathbf{\Gamma}_k\right]\frac{\nabla \mathcal{L}(\boldsymbol{\theta}_k)}{\prod_{m=0}^{k-1}(1-\varepsilon_{m})},
\end{equation}
 where $\mathbf{\Gamma}_k=\sum_{m=0}^{k-1}\gamma_m\prod_{n=0}^m(1-\varepsilon_{n-1})$ is the matrix of corrections to the metric picked up by the innovations from the first $k-1$ steps. 
We have that $\prod_{k=0}^n(1-\frac{\varepsilon_0}{k+1})=\frac{(1-\varepsilon_0)}{\Gamma(2-\varepsilon_0)}\cdot\frac{\Gamma(n+2-\varepsilon_0)}{\Gamma(n+2)}$ and that $\frac{\Gamma(n+2-\varepsilon_0)}{\Gamma(n+2)}\sim n^{-\varepsilon_0}$ as $n\rightarrow\infty$. Since $\varepsilon_k=\varepsilon_0/(k+1)$, the innovations are attenuated  $\propto k^{-1}$, and, for some number of steps $k'\gg1$, the innovations can be considered negligible. In this regime, where $k>k'$, the step taken is $\Delta_k \propto (k-1)^{\varepsilon_0}\left[\mathbf{B}_0^{-1}-\mathbf{\Gamma}_{k'}\right]\nabla \mathcal{L}(\boldsymbol{\theta}_k)$, where the approximate metric $\mathbf{B}_0^{-1}-\mathbf{\Gamma}_{k'}$ can be considered constant. 

This behavior invites a possible modification to the algorithms, where, if convergence has not been achieved after $k'$ steps, the metric is reinitialized at the current parameters by computing the full FIM matrix at $\boldsymbol{\theta}_{k'}$ and the algorithm restarted.

\section{Connection of VarQITE and QNG}\label{app:connection_of_varqite_and_qng}

As stated in the main text, there is a close relationship between the QFIM, $\mbf F$, and the Fubini-Study metric, $\mbf A$, which is given by
\begin{equation} 
	A_{ij} = \mathrm{Re} \left\{\left\langle\partial_{\theta_i} \Phi \vert \partial_{\theta_j} \Phi\right\rangle
	- \left\langle\partial_{\theta_i}\Phi\vert\Phi\right\rangle \left\langle\Phi \vert\partial_{\theta_j}\Phi\right\rangle\right\},
	\end{equation}\\
where, $\partial_{\theta_i} \equiv \frac{\partial}{\partial_{\theta_i}}$.
The Fubini-Study metric~\cite{Fubini1908, Study1905, Yao2022, Wilczek1989, Hackl2020}, is the metric 
of parametrized \textbf{pure} quantum states $\ket{\Phi(\boldsymbol{\theta})}$.
$\mbf A$ can be expressed as 
the real part of a more general quantum geometric tensor (QGT)~\cite{Campos2007, Wilczek1989, Bukov2019, Kolodrubetz2017Geometrynonadiabaticresponse}
\begin{equation}
	G_{ij} = \left\langle\partial_{\theta_i} \Phi \vert \partial_{\theta_j} \Phi\right\rangle
	- \left\langle\partial_{\theta_i}\Phi\vert\Phi\right\rangle \left\langle\Phi \vert\partial_{\theta_j}\Phi\right\rangle \label{eq:qgt}, 
\end{equation}
whose imaginary part corresponds to the Berry geometrical phase~\cite{Pancharatnam1956, Berry1984, Facchi2010, Wilczek1989}.

For pure states -- as we consider exclusively in this work -- the 
Fubini-Study metric (in matrix form) is (up to a factor of 4) equivalent to the QFIM~\cite{zhoujiang2019, Liu2020QuantumFisherinformation, Giovannetti2011, Petz1996},  i.e.,
$ \mbf F =  4 \mbf A$. 
The factor of 4 could, however, be absorbed by a change of variables~\cite{Facchi2010} or in the time-step $\delta\tau = \frac{\eta}{4}$ as we did in the main text.
Thus we use the terms Fubini-study metric/QFIM and variables $\mbf A$ and $\mbf F$ interchangeably in the main text.

The matrices $\mbf A$  and $\mbf F$ describe the geometry of the parameter space rather than the energy landscape. 
The second term of \eqref{eq:fisher} 
resolves a possible arbitrary overall phase mismatch between $\ket{\Phi(\boldsymbol{\theta}(\tau))}$ and the target state $\ket{\Psi(\tau)}$ along the imaginary time propagation~\cite{Zoufal2021, Yuan2019Theoryvariationalquantum}.
Using different variational principles (time-dependent/Dirac-Frenkel)~\cite{Yuan2019Theoryvariationalquantum, Broeckhove_1988} yields slightly different equations for the metric and gradient resulting in possibly complex values of the parameters $\boldsymbol{\theta}$ (see Ref. number~\cite{Yuan2019Theoryvariationalquantum} for details).
As $\boldsymbol{\theta}$ usually refers to real-valued angles of rotational gates in a PQC, solving Eq.~(3) from the main text using McLachlan's variational principle is preferred in the VarQITE setting, as it ensures real-valued solutions for $\frac{\partial \boldsymbol{\theta}}{\partial\tau}$. 
If $\ket{\Phi}$ and $\partial_{\theta_i}\ket{\Phi}$ are real (not to be confused with real parameters), the second term in \eqref{eq:fisher} vanishes, due to the normalization of $\ket{\Phi},~\braket{\Phi\vert\Phi} = 1$
\begin{align}
	\left\langle \partial_{\theta_i}\Phi \vert \Phi\right\rangle + \left\langle\Phi \vert \partial_{\theta_i}\Phi\right\rangle = \partial_{\theta_i} 1 = 0.
\end{align}

Due to the above-mentioned relation between the Fubini-Study metric and QFIM, $\mbf F=4\mbf A$, Eq.~(6) from the main text reveals that QNG is equivalent to VarQITE when the energy of the system is used as the cost function,  $\mathcal{L} = \braket{\hat H}$, and $\eta = 4\delta\tau$. 

Additionally, VarQITE is closely related to the stochastic reconfiguration (SR) method of Sorella~\cite{Sorella1998, Sorella2000,Mazzola2012}, which is a second-order iterative approximation to the \enquote{classical} ITE.

\textcolor{black}{
\section{Ensuring Positive Definiteness of the Quantum Fisher Information Matrix}
\label{sec:PSD_QFIM}
In the presence of noise, especially shot noise, the method used to estimate the QFIM may produce a matrix that is not positive semi-definite. This is problematic as it could adversely affect the optimization process, potentially leading to unstable or divergent behavior. To mitigate this issue, we employ a diagonal loading to ensure that the QFIM remains positive definite (PD).
The method is straightforward but crucial for the robustness of our optimization algorithms. We first compute the eigenvalues of the QFIM. If the matrix has any negative eigenvalues, we identify the most negative one, say $\lambda_{\text{min}}$. We then add $(\gamma_{\rm reg} - \lambda_{\text{min}}) $ times the identity matrix to the QFIM, where $\gamma_{\rm reg}$ is a small regularising parameter. Mathematically, this can be expressed as:
\[
\mathbf{F}_{\text{PD}} = 
\begin{cases} 
\mathbf{F} + (\gamma_{\rm reg} - \lambda_{\text{min}}) \mathbb{1} & \text{if } \lambda_{\text{min}} < 0, \\
\mathbf{F} & \text{otherwise}.
\end{cases}
\]
Here, \( \mathbf{F} \) is the original QFIM and \( \mathbb{1} \) is the identity matrix of the same dimension as \( \mathbf{F} \).
}

\end{document}